\documentclass[letterpaper,fleqn,usenatbib]{mnras}

\usepackage{enumerate}

\def\eg{{\it e.g.}}
\def\etal{{\it et al.}}
\def\etc{{\it etc.}}
\def\ie{{\it i.e.}}

\def\pmb#1{\setbox0=\hbox{$#1$}%
  \kern-0.25em\copy0\kern-\wd0
  \kern.05em\copy0\kern-\wd0
  \kern-0.025em\raise.0433em\box0}
\def\spmb#1{\setbox1=\hbox{${\scriptstyle #1}$}%
  \kern-0.25em\copy1\kern-\wd1
  \kern.05em\copy1\kern-\wd1
  \kern-0.025em\raise.0433em\box1}

\long\def\Ignore#1{\relax}
\usepackage{color}
\definecolor{red}{rgb}{0.7,0.1,0.1}
\definecolor{blue}{rgb}{0.2,0.2,0.8}
\definecolor{green}{rgb}{0.1,0.6,0.1}

\usepackage{graphicx}	

\title[Recurring spiral instabilities]{Spiral instabilities: Mechanism for recurrence}

\author[Sellwood \& Carlberg]
          {J. A. Sellwood,$^{1}$\thanks{E-mail:sellwood@as.arizona.edu}
and
{Ray G. Carlberg,$^2$\thanks{E-mail: raymond.carlberg@utoronto.ca}}
\\
$^1$Steward Observatory, University of Arizona, 933 N Cherry Ave,
Tucson AZ 85722, USA \\ $^2$Department of Astronomy and Astrophysics,
University of Toronto, ON M5S 3H4, Canada}

\pubyear{2019}

\begin{document}
\label{firstpage}
\pagerange{\pageref{firstpage}--\pageref{lastpage}}
\maketitle

\begin{abstract}
We argue that self-excited instabilities are the cause of spiral
patterns in simulations of unperturbed stellar discs.  In previous
papers, we have found that spiral patterns were caused by a few
concurrent waves, which we claimed were modes.  The superposition of a
few steadily rotating waves inevitably causes the appearance of the
disc to change continuously, and creates the kind of shearing spiral
patterns that have been widely reported.  Although we have found that
individual modes last for relatively few rotations, spiral activity
persists because fresh instabilities appear, which we suspected were
excited by the changes to the disc caused by previous disturbances.
Here we confirm our suspicion by demonstrating that scattering at
either of the Lindblad resonances seeds a new groove-type instability.
With this logical gap closed, our understanding of the behaviour in
the simulations is almost complete.  We believe that our robust
mechanism is a major cause of spiral patterns in the old stellar discs
of galaxies, including the Milky Way where we have previously reported
evidence for resonance scattering in the recently released {\it
  Gaia\/} data.
\end{abstract}



\section{Introduction}
\label{sec.intro}
Theorists have long dreamed that the spiral patterns gracing most disc
galaxies, and simulations thereof, reflect the normal modes of the
disc.  That is, they are mildly non-linear manifestations of
self-excited, linear instabilities of the stellar disc that are
uniformly rotating and exponentially growing density waves.  Long ago,
this grand program was set back by two early discoveries from normal
mode analyses of apparently reasonable models of featureless discs:
models where the rotation curve rose gently from the centre were
dominated by bar-forming instabilities \citep{Ho71, Ka78}, while
models having a dense (bulge-like) centre had no modes whatsoever
\citep{To81}.

Here we begin by reviewing why we think spiral modes are in fact the
most promising mechansim, despite this early setback, and add to
our case by demonstrating how spiral modes recur.

\subsection{Modes in galactic discs}
\label{sec.modes}
A normal mode of any system is a self-sustaining, sinusoidal
disturbance of fixed frequency and constant shape, save for a possible
uniform rotation; the frequency would be complex if the mode were to
grow or decay.  In the case of galaxy discs, the perturbed surface
density of a mode is the real part of
\begin{equation}
\delta\Sigma(R,\phi,t) = A_m(R)e^{i(m\phi - \omega t)},
\label{eq.mode}
\end{equation}
where $m$ is the angular periodicity, $\omega = m\Omega_p + i\beta$,
$\Omega_p$ is the angular rate of rotation, usually called the pattern
speed, and $\beta$ is the growth rate.  The complex function $A_m(R)$,
which is independent of time, describes the radial variation of
amplitude and phase of the mode.

Stability analysis of a system supposes small amplitude perturbations
about the equilibrium state and is linearized when any terms that
involve products of small amplitude terms are discarded -- see
\citet{Ka71} for a careful formulation.  The equilibrium is linearly
unstable if any of the resulting normal modes have a positive growth
rate, since its amplitude will exponentiate out of the noise until
the neglected 2nd and higher order terms become no longer
negligible.\footnote{The word instability is sometimes used to imply a
  purely imaginary frequency, and a mode with a complex frequency is
  then described as an overstability.  Here we adopt the more usual
  convention of simply describing all frequencies with $\beta>0$ as
  instabilities.}

Normal modes can be standing wave oscillations of the system that
exist between two reflecting barriers, as in organ pipes and guitar
strings, which are generally described as cavity modes in galaxy
discs.  The prime example in galaxies is the bar-forming mode, for
which reflections take place at the centre and at corotation
\citep{To81, BT08}.  An instability of this type is possible only if
the disturbance has no inner Lindblad resonance (hereafter ILR), since
linear theory \citep{Ma74} predicts that any disturbance that
encounters an ILR will be absorbed, and therefore damped.  An ILR must
be present for any reasonable pattern speed when the centre is dense,
and therefore no small-amplitude cavity mode is possible in a
featureless disc of this kind.  But $\Omega_c-\kappa/2$ has a maximum
value in mass models having quasi-harmonic cores, and bar-forming
instabilities avoid resonance damping by having pattern speeds that
exceed this maximum.  Here $\Omega_c(R)$ is the angular frequency of
circular motion at radius $R$ in the disc mid-plane, and $\kappa(R)$
is the usual frequency of small-amplitude radial oscillations about a
circular orbit \citep{BT08}.  The dominant mode of several
bar-unstable models has been identified in simulations, with excellent
quantitative agreement of both the frequency and mode shape
\citep{SA86, ES95}.  Other types of cavity mode have been proposed by
\citet{Ma77} and by \citet{SC14} and are described below.

Galactic discs can also support another class of mode.  The best known
examples are edge modes \citep{To81, PL89} and groove modes
\citep{SL89, SK91}.  They are not standing waves, however, and
maintain a fixed shape and frequency by other means.  In the case of
the edge mode, a small non-axisymmetric distortion of the disc where
the density decreases steeply, moves high density material out to
places where the equilibrium density was lower, and conversely at
other azimuthal phases.  On their own, such co-orbiting distortions
would be neutrally stable and therefore of no interest.  But as
\citet{JT66} taught, a cool surrounding disc responds vigorously to a
co-orbiting mass excess, creating a trailing wake that extends
radially far into the shear flow on either side of the perturbing
mass; this behaviour is a consequence of swing amplification
\citep{GLB, JT66, To81, Bi19}.  The distorted edge therefore excites a
strong supporting response from the interior disc that is not balanced
by the exterior response because the equilibrium density drops rapidly
with radius at the edge.  The attraction of the interior wake on the
original density excess increases its angular momentum, causing it to
grow exponentially as it rotates.  The necessary conditions for
instability were summarized by \citet{To89}.

A groove in a disc is effectively two closely spaced edges, which
however give rise to a single mode because the distortions on each
edge are gravitationally coupled.  Note that it is a steep gradient in
the angular momentum density that matters for both edge and groove
modes; epicyclic blurring can mask the steepness in the actual surface
density profile.  As for edge modes, it is the supporting response of
the surrounding disc that causes the groove mode to have a substantial
radial extent and to grow rapidly \citep{SK91}, and these authors were
able to obtain good quantitative agreement between their analytic
predictions and simulations.

Note that swing amplication, famously illustrated in the dust-to-ashes
figure of \citet{To81}, is not a mode both because the shape changes
with time and the amplitude variation is not a simple exponential.
Also the wake response to an imposed co-orbiting mass clump
\citep{JT66} is not a mode because, to first order, it would disperse
if the clump were removed, and it therefore is not self-sustaining.
Both are simply responses of the disc to externally imposed
disturbances.  However, they are both very helpful concepts when
trying to understand the mechanisms of self-sustaining modes.

\subsection{Noise, resonances, and heating}
\label{sec.heating}
The collisionless fluid of stars invoked in these analytical
treatments is an idealization that assumes the stars to be infinitely
finely divided so that phase space is smooth.  The numbers of stars in
galaxy discs is large enough that this assumption holds quite well
\citep[see][for caveats]{Se14a}, but galaxies contain mass clumps such
as star clusters and giant molecular clouds, and the number of
particles employed in a simulation is generally several orders of
magnitude fewer than the number of stars.  Thus shot noise in both
real galaxies and in simulations gives rise to significant
inhomogeneities in the disc.

The spectrum of shot noise in a shearing distribution of randomly
distributed gravitating masses inevitably contains leading wave
components that will be strongly amplified as the shear carries them
from leading to trailing.  This behaviour has two important
consequences.  The first consequence of swing-amplified shot noise is
that each heavy particle develops a wake \citep{JT66, Bi19}.  The two
point correlation function of the particles becomes greater along the
direction of the wake and lower in other directions, and the particle
distribution is said to be polarized.  Since the distribution of
particles is no longer perfectly random, the amplitude of all
components of the noise spectrum is enhanced, causing subsequent
noise-induced fluctuations to be stronger, although linear theory
predicts this cycle should asymptote in a few epicycle periods to a
mean steady excess over the level expected from uncorrelated noise
\citep{JT66, TK91}.

Second, the collective amplified response of any one component of the
noise orbiting at the angular frequency $\Omega_p$, creates a coherent
trailing wave in the disc that propagates away from corotation
\citep{To69, To81} until it reaches a Lindblad resonance where it is
absorbed \citep{Ma74}.  For near circular orbits the Lindblad
resonances occur where the Doppler shifted frequency at which the
stars encounter the wave is equal to the epicyclic frequency of their
small radial excursions, \ie, the radii at which
\begin{equation}
m\left[\Omega_p - \Omega_c(R)\right] = l\kappa(R),
\label{eq.resonance}
\end{equation}
with $l=\pm1$ at Lindblad resonances.  The negative sign is for the
ILR, where stars overtake the wave, and the positive is for the OLR
where the wave overtakes the stars, at the local epicycle frequency in
both cases.  Wave-particle interactions at the resonance cause
localized irreversible changes to the energy and angular momenta of
stars.  Jacobi's invariant \citep{BT08} implies that changes are
related as $\Delta E = \Omega_p \Delta L_z$.  On average and to second
order, particles lose $L_z$ at the ILR and gain at the OLR \citep{LBK,
  CS85}, and this outward transfer of $L_z$ allows the wave to extract
free energy from the galactic potential enabling the scattered
particles to acquire additional random energy at both resonances.  The
resulting depopulation of stars on near circular orbits over the
narrow region of each resonance creates a ``scratch'' in the disc that
may alter its stability properties.

It is important to realize that linear theory neglects this second
order effect by assumption, \ie\ it does not allow for changes to the
equilibrium state.  In fact, \citet{Se12} found that the amplitudes of
successive episodes of uncorrelated swing amplified noise rose slowly,
but continuously, because the consequent scratches to the disc from
each episode caused partial reflections of subsequent disturbances
that allowed further amplification \citep{SC14, FP15}.  This process
continued until the the partial reflections became strong enough that
the disc was able to support an unstable mode \citep{Se12, DFP19}, and
coherent growth to large amplitude began.  We discuss this behaviour
further in \S\ref{sec.argue}.

Here we have used the word ``scratch'' to describe quite mild changes
to the distribution function (hereafter DF) that can cause partial
reflections of a wave propagating radially within the disc.  But
scattering at a Lindblad resonance could also carve a similar feature
that seeds a groove mode instead, and we will show below that this
appears to be the more usual behaviour.

Note that the Lindblad resonances are closer to corotation, where
$\Omega_p = \Omega_c(R_{\rm CR})$ ($l=0$ in eq.~\ref{eq.resonance}),
when $m$ is large than for waves of lower $m$, implying that more free
energy can be extracted from the potential, causing more rapid heating
when the disc supports larger-scale waves.  The value of $m$ that is
amplified most strongly \citep{JT66, To81} is 
\begin{equation}
m \approx {R_{\rm CR}\kappa^2 \over 2\pi X G\Sigma},
\label{eq.prefm}
\end{equation}
with $1 < X < 2$ in a flat rotation curve.  Thus the preferred $m$
varies inversely with the disc surface density $\Sigma$ \citep{SC84,
  ABP}.  Whatever the origin of disturbances, a sub-maximal disc will
prefer higher $m$, and therefore heat more slowly than would heavier
discs.

Since spiral activity heats collisionless particles, or stars, an
uncooled disc must become less able to support collective disturbances
over time, as is well known.  \citet{SC84} established that dynamical
cooling by gas and star formation is needed to counter secular heating
and to maintain spiral activity, a result that has been confirmed in
many more recent simulations \citep[\eg][]{Ro08, Ag11, Au16}.  It
nicely accounts for the observation \citep[\eg][]{Oo62} that almost
all spiral patterns are seen in galaxies that contain gas and are
forming stars.

\subsection{The spiral challenge}
\label{sec.theories}
As already noted, a galaxy model having a dense centre and no sharp
features, such as an edge or a groove, should not support any normal
modes at all.  Since spirals are ubiquitous in disc galaxies
containing a modest gas fraction, and also develop spontaneously in
simulations of isolated discs, some mechanism is needed to excite
them.  Currently, there are at least four proposed mechanisms:
\begin{enumerate}[i]
\item Following \citet{Ma77}, \citet{BL96} suggest that spirals result
  from a cavity type mode in a low-mass disc that is dynamically cool
  over most of the disc, but which also possess an inner ``$Q$
  barrier'' to shield the ILR.  These authors imagine that most
  galaxies support a single, long-lived, mildly-unstable mode that
  persists for many tens of galactic rotations and becomes
  ``quasi-steady'' due to dissipative shocks in the gas, but allow
  that superposition of a second mode may be needed in some cases.  As
  given by eq.~(\ref{eq.prefm}), strong swing-amplification for $X
  \sim 2$ sets a preference for multi-arm instabilities over 2-arm
  modes in low mass discs.  By considering only bi-symmetric
  disturbances in low-mass discs, \citet{BLLT} exploited the mild
  amplification when $X > 3$ in order to obtain slowly-growing spiral
  modes in their global stability analysis of many galaxy models.

  Simulations by \citet{Se11} of one of the cases \hbox{presented} by
  \citet{BLLT} confirmed that a \hbox{single}, slowly-growing mode was
  present when disturbance forces were restricted to $m=2$.  Not
  surprisingly, \hbox{however}, he also found much more vigorous
  instabilities appeared when higher sectoral harmonics contributed to
  disturbance forces, and the contrived basic state of the disc that
  was designed to support the $m=2$ mode was rapidly changed by these
  more vigorous \hbox{instabilities}.  This evidence alone should have ruled
  out the \hbox{theory}, \hbox{although} \citet{Sh16} ignored it as he
  continued to \hbox{advocate} for their picture.

\bigskip
\item \citet{To90} and \citet{TK91} abandoned the idea of spirals as
  normal modes, and advocated instead that a collection of massive
  clumps in the disc, each of which becomes dressed with its own wake,
  would create a ``kaleidoscope'' of shearing spiral patterns.  Their
  simulations of this process were all confined to a single shearing
  patch with a modest number of particles.  \citet{DVH} conducted
  global simulations of a low-mass disc, embedded in a rigid halo,
  employing $10^8$ star particles to which they added a sprinkling of
  heavy particles that induced evolving multi-arm spiral patterns in
  the stars.  In a separate experiment they also tried a single
  perturber of mass $10^7\;$M$_\odot$, which they removed again after
  it had completed one orbit; the simulation continued to manifest
  spiral activity in response to the non-axisymmetric density
  distribution created by the original imposed mass, which they
  described as non-linear behaviour.

  These authors suggest that ``ragged'' spiral activity in galaxies
  results from such responses to co-orbiting giant molecular clouds,
  massive star clusters, \etc, and to the lingering disc responses
  should any disperse.  Although we do not doubt their numerical
  results, we remain unconvinced of their importance for spiral
  activity in galaxies.  First, the $10^7\;$M$_\odot$ particle that
  \cite{DVH} employed produced only a modest, and not very extensive
  wake in their low-mass disc.  Spirals in real galaxies generally
  have greater amplitude and radial extent than this, suggesting that
  yet more massive clumps would be needed.  Second, a collection of
  randomly placed heavy particles does not seem likely to produce a
  net response that is predominantly 2- or 3-armed, as are observed in
  the overwhelming majority of galaxies \citep{Davi12, Hart16, Yu18}.
  Third, clumps massive and numerous enough that their associated
  wakes produce large-amplitude and radially-extensive spiral patterns
  would scatter disc stars, and heat the disc rapidly so that the
  responses would fade quickly unless the disc were cooled
  aggressively, and the necessary cooling \citep{To90} seems rather
  extreme.  A fourth, and overriding, reason is that the idea is
  unnecessary, because discs readily support unstable spiral modes, as
  we discuss next.

\bigskip
\item \citet{SC14} demonstrated that spiral activity in simulations
  resulted from superposition of a number of coherent, uniformly
  rotating waves.  They found that each wave grew and decayed, but was
  detectable over a period of some ten rotations at its corotation
  radius, and they presented substantial evidence that the waves were
  in fact modes.  The vigorous modes in their new picture differ
  substantially from those invoked by \citet{BL96} because they work
  best where swing-amplication is strongest, do not last for nearly
  as long, and fresh instabilities develop to maintain spiral
  activity.

  Note that the spiral appearance changes still more rapidly than do
  the modes in their simulations.  This is because the superposition
  of several modes, each having a different pattern speed and perhaps
  also angular periodicity, as well as time varying amplitude, causes
  the pattern of visible spiral arms to change radically in less than
  one orbit.\footnote{An animation showing the time evolution of the
    net density when two notional patterns are superposed is at \hfil
    \\ {\tt
      http://www.physics.rutgers.edu/$\sim$sellwood/spirals.html}}

  \citet{SC14} also postulated a new cavity mode that relied upon
  partial reflection of travelling waves off an impedance variation
  created by a previous perturbation within the disc.  They suggested
  that ILR scattering by a past wave would have ``scratched'' an
  originally smooth distribution function (see \S\ref{sec.heating}) to
  create a deficiency of stars on near circular orbits at some radius.
  A subsequent ingoing trailing wave encountering the scratch before
  reaching its own ILR, would find that the scratch in the otherwise
  smooth disc represents an abrupt change of impedance, and is
  therefore partially reflected into a outgoing leading wave.  A
  second reflection of the wave by swing amplification at corotation
  creates a cavity that will support standing waves having frequencies
  allowed by the usual phase closure condition.  They dubbed this new
  type of cavity mode a ``mirror mode.''  As for the bar mode that
  also includes swing amplification from leading to trailing, a
  reasonable reflected fraction at the partial mirror will cause a
  mirror mode to grow rapidly, and the instability must run its course
  within a modest number of pattern rotations.  Nevertheless, spiral
  activity can be maintained because other modes develop in rapid
  succession.  We present evidence in \S\ref{sec.argue} below that the
  first real instability in the tests of the Mestel discs reported in
  \cite{Se12} was almost certainly of this type.  Some mirror modes
  may also have been present in the simulations of \citet{SC14}, but
  we here (\S\ref{sec.results}) argue for a more probable recurrence
  cycle.

\bigskip
\item \citet{Gran12a, Gran12b}, \citet{Baba13}, \citet{Roca13}, and
  others present simulations of mostly sub-maximal discs and report
  that spiral patterns are shearing structures, in which the density
  maxima wind as would material arms, or nearly so. \citet{Gran12a,
    Gran12b}, \citet{Ka14}, and other authors have described spiral
  arm streaming motions that are consistent with expectations set out
  by \citet{Ka73}; however, such a flow pattern is required for any
  self-consistent spiral, irrespective of its nature or origin.
  \citet{Baba13} and \citet{MK18} find that spirals in their
  simulations behave as predicted by swing amplification theory
  \citep{To81}.  \citet{Gran13} reported a correlation between the
  mean pitch angle of spiral arms in their simulations and the shear
  rate in the disc, which they also suggested was consistent with
  swing amplification theory.  \citet{Baba15} reported similar
  shearing and evolving spirals in the outer discs of barred
  simulations.  \citet{KN16} excited regular spiral patterns that were
  particle wakes in the disc forced by a few heavy particles equally
  spaced around rings, finding additional swing amplification as the
  wakes lined up.  The earlier part of this body of work was
  summarized in the review by \citet{DB14}, and all these authors
  propose that shearing patterns are the fundamental character of
  spiral arms.  We discuss these findings in \S\ref{sec.argue}.

\end{enumerate}

The {\it Gaia\/} satellite has revealed the local phase space
structure of the Milky Way in unprecedented detail \citep{Gaia2}.
\citet{Se19} used these data to try to discriminate among the
different theories just described.  They calculated the changes to a
smooth distribution function that would be caused by a single episode
of each of the last three spiral models and compared the predictions
with the {\it Gaia\/} data.  They concluded that the features in
action space seemed more consistent with the transient spiral mode
model \citep{SC14} than with either of the other two.  Note that
\citet{Mo19} found evidence in the same data for scattering by the
Milky Way bar, but some features remained that they could not
attribute to the bar.  However, \citet{Hu18} concluded that other
features in the same data were consistent with shearing spiral arms.

\citet{Se19} also argued that the long-lived spiral mode model
advocated by \citet{BL96} would not cause any pronounced changes,
because the ILR, where the largest changes generally occur, is
shielded by the $Q$ barrier, no net change is expected at corotation,
and the OLR would probably lie too far outside the solar circle to
affect the local distribution.  Thus if their delicate mechanism for
spiral generation were indeed to operate in the Milky Way, it would
contribute little to the observd extensive sub-structure in phase
space and would implausibly have to survive in the mild disequilibrium
state of the disc revealed in the {\it Gaia} data.

\subsection{Modes or shearing patterns?}
\label{sec.argue}
The shearing spiral behaviour that is apparent in almost all
simulations of cool, isolated discs possesses many aspects of
swing-amplification and wakes, as the above cited papers have
reported.  We have also reported such behaviour: \eg, Fig.~3 of
\citet{SC84} indicated the time evolution of a single 3-arm spiral
disturbance that apparently sheared and amplified to an open trailing
pattern before decaying as it continued to wind and, in simulations of
much improved numerical quality, \citet[][their Fig.~3]{SC14}
illustrated constantly changing patterns, which is qualitatively
similar to the behaviour that all authors report.

However, we have also found \citep[\eg][and later in this paper]{Se89,
  Se11, SC14} that the constantly changing appearance of spiral
patterns results from the superposition of a modest number of longer
lived waves \citep[see also][]{Qu11}.  Even though a mode
(eq.~\ref{eq.mode}) has a fixed pattern speed at all radii and a
constant shape function, $A_m(R)$, shearing patterns are the
inevitable consequence of superposed modes.  All that is required to
produce the appearance of a shearing transient spiral (see footnote 2)
from two or more superposed patterns of fixed shape is that the modes
closer to the centre have higher pattern speed, which is always true.

The proposition that the shearing patterns are the fundamental
behaviour is not a satisfactory explanation for the origin of the
spirals, because it is unable to answer a key question: how can the
spiral amplitudes in simulations be largely independent of the number
of particles?  No recent paper that argues for this interpretation
addresses this issue, yet there is substantial evidence to support it:
\citet{Se11}, \citet{Se12}, and \citet{SC14} all reported similar
final amplitudes in experiments in which the number of particles
ranged over orders of magnitude.  As $N$ is increased, the amplitude
of shot noise fluctuations must decrease as $N^{-1/2}$, and some kind
of growth mechanism is required to produce final amplitudes that are
independent of $N$.  Swing-amplification and/or wakes do not lead to
indefinite growth, and therefore cannot account for final amplitudes
that are independent of $N$, at least in linear theory.  We would
agree with an argument that non-linear scattering
(\S\ref{sec.heating}) is important, but that plays into our case for
unstable modes \citep[][and this paper]{SC14}.

It is likely that most models adopted in simulations are unstable,
because they allow some feedback through the centre and/or possess
outer edges sharp enough to excite normal modes.  Mildly unstable
modes will be seeded at low amplitude when $N$ is large and may not
cause visible changes for several rotations.  Both \citet{Se11} and
\citet{SC14} present cases in which the first visible features took
longer and longer to appear as the number of particles was
increased.\footnote{Pronouncements of stability after a short period
  of evolution in large-$N$ models \citep[\eg][]{DVH} are unlikely to
  hold in longer integrations.}  Once the first instability has
appeared, we find that activity takes off, as the particle
distribution becomes more and more structured by the previous
evolution; the new results presented in \S\ref{sec.results} below show
in more detail how a recurrent cycle of modes can occur.

\begin{figure}
\includegraphics[width=.9\hsize,angle=0]{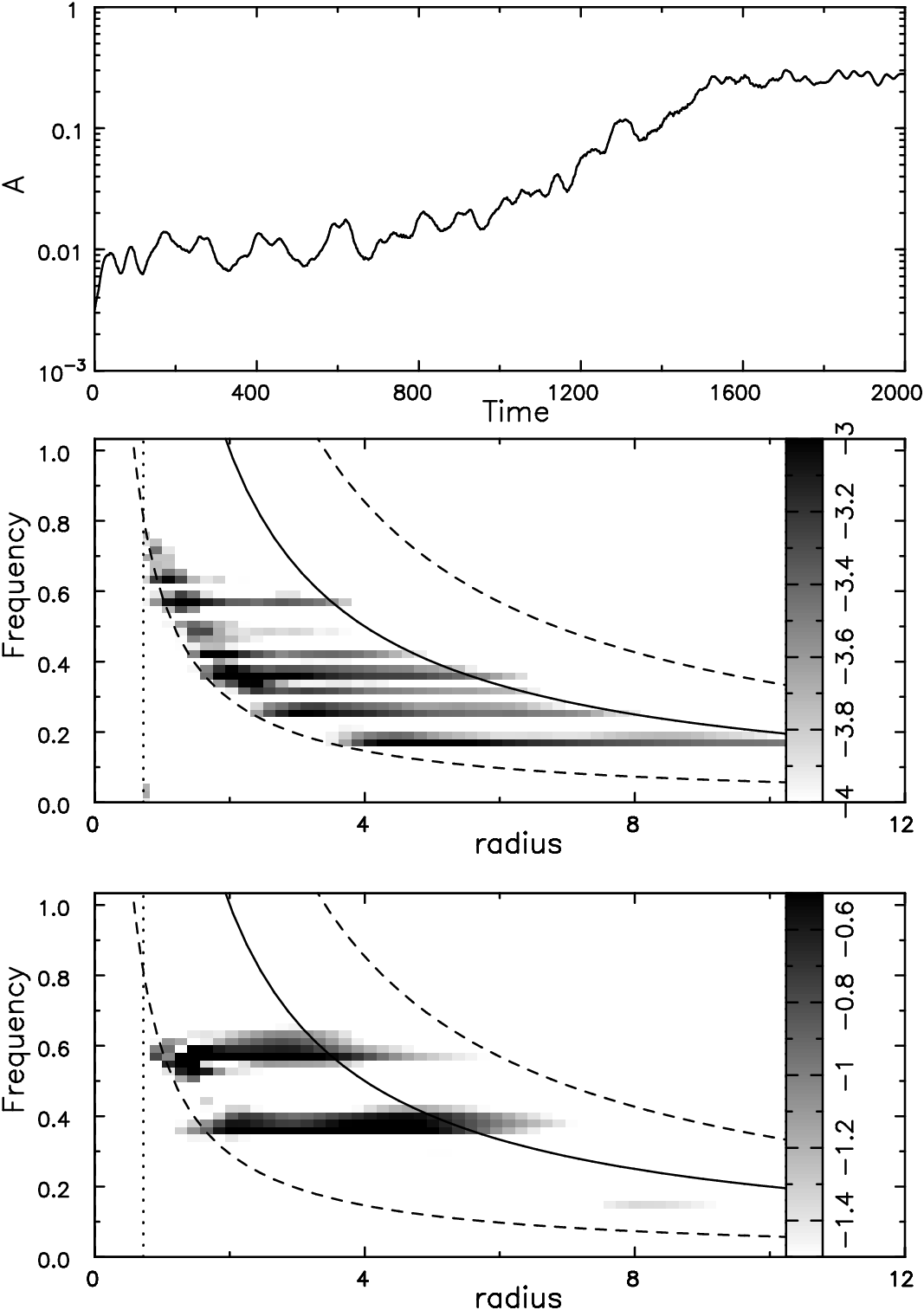}
\caption{An old simulation of the half-mass Mestel disc.  The top
  panel shows the amplitude evolution of $m=2$ disturbances reproduced
  from the $N=5\times 10^7$ case in Fig.~2 of \citet{Se12}.  The power
  spectrum over the time range $100\leq t \leq 400$ in the middle
  panel shows no coherent modes were present over this interval.  The
  solid curve indicates the radial variation of $m\Omega_c$ while the
  dashed curves show $m\Omega_c \pm \kappa$.  The bottom panel is for
  the period $1300\leq t \leq 1600$ and reveals that the more rapid
  amplitude growth from $t=1000$ is caused by more coherent
  instabilities, dark horizontal streaks, that developed later in this
  model, as \citet{Se12} reported.  Note that the grey scale
  amplitudes are logarithmic and differ in the two panels.}
\label{fig.nogroove}
\end{figure}

However, it is not neccessary to start from a disc that possesses one
or more global instabilities.  \citet{To81} claimed that the half-mass
Mestel disc is linearly stable, and this case was studied in
simulations by \citet{Se12}.  Fig.~\ref{fig.nogroove} illustrates the
evolution of the $N = 5 \times 10^7$ simulation from his paper;
details of the model and the numerical method are given in
\S\ref{sec.methods} below.  The top panel reproduces the cyan curve
from Fig.~2 of \citet{Se12}, which reported at each instant the
greatest value of the ratio $\Sigma_2(R)/\Sigma_0(R)$ within the
radius range $1.2 < R < 12$, where $\Sigma_2$ is the amplitude of the
bi-symmetric disturbance density.  After an initial surge by a factor
of a few as each particle created its own wake (\S\ref{sec.heating}),
these $m=2$ features manifested slow secular growth.  During this
phase, the amplitudes of successive swing-amplified episodes increased
slowly because each created mild scratches in the previously smooth
disc that enabled partial reflections.  The partial mirrors so created
at first reflected small fractions of the incident waves, but as the
disturbance amplitudes rose with each episode, the new scratches
became more reflective.  No coherent modes were detectable during this
secular growth phase, and the power spectrum, such as shown in the
middle panel, was characterized by multiple uncorrelated frequencies
arising from whichever noise features were strongest at the time; the
corotation radius of each event determines the frequency of the wave
that propagates to its ILR where it is absorbed.  These findings were
consistent with Toomre's linear theory prediction of global stability,
since the secular growth was caused by the non-linear scattering terms
that his analysis neglected.  The largest amplitude disturbances
reached an overdensity of $\sim 2$\% by $t \sim 1000$, at which point
the reflections became strong enough to cause true unstable modes in
this originally stable disc.  The amplitude in the top panel of
Fig.~\ref{fig.nogroove} began to rise more rapidly and the power
spectrum, in the bottom panel, that has a different amplitude scale
shows much stronger and more coherent waves, which are modes.

\begin{figure}
\includegraphics[width=.9\hsize,angle=0]{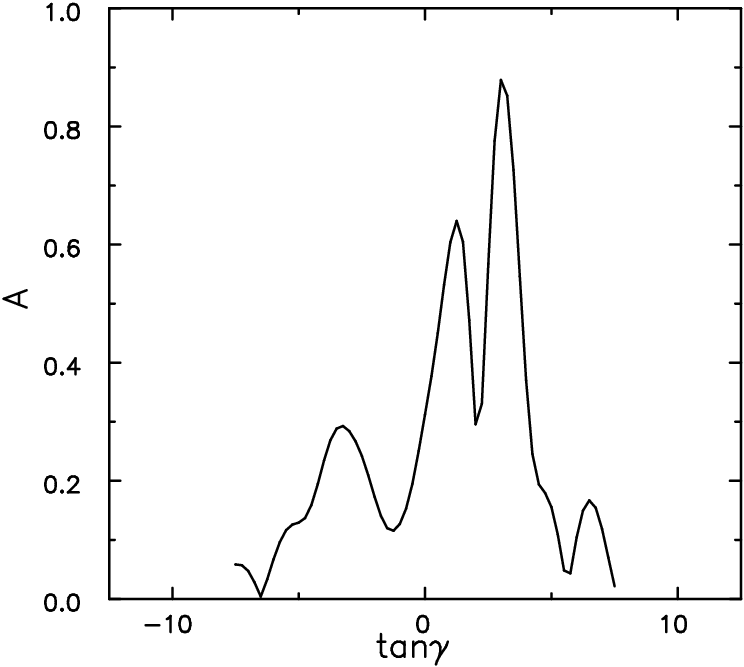}
\caption{Fig.~6 of \citet{Se12} showed the shape of the mode fitted to
  the data from his model 50c.  Here we show the logarithmic spiral
  transform of that mode density.  Positive values of $\tan\gamma$ are
  trailing, negative are leading, where $\gamma$ is the angle between
  the tangent to the spiral and the radius vector.  The amplitude
  scale is arbitrary.}
\label{fig.mode4169}
\end{figure}

\citet{FP15} and \citet{FBP15} successfully developed a quasi-linear
theoretical description of this behaviour.  \citet{DFP19} conducted a
global mode analysis of this model at $t=1400$, and estimated the
frequency of the dominant instability to be $\omega = 0.597 + 0.013i$
in excellent agreement with Sellwood's own estimate (his paper has
factor 10 typo in the growth rate).  Although both analyses were of
the model denoted 50c, in which the particles coordinates had been
scrambled at $t=1400$, the faster wave in the bottom panel of
Fig.~\ref{fig.nogroove} confirms a mode of similar pattern speed when
the evolution is uninterrupted.  Fig.~\ref{fig.mode4169} shows the
logarithmic spiral decomposition of the mode \citet{Se12} fitted to
the data from his model 50c.  The leading/trailing bias reflects the
trailing appearance of the mode, but the leading side has significant
amplitude indicating that it was a mirror, or cavity, mode that
operated by an inner reflection off the partial mirror to a leading
wave, as \citet{SC14} described.

\citet{Sr19} calculates the changes to the DF at the principal
resonances caused by a transient disturbance, and attempts to show how
they lead to further instabilities.  However, the correspondence
between the modes he calculates and the groove and mirror modes that
we have reported is not immediately apparent from his analysis.

Thus we consider that every simulation of a cool, unperturbed disc
must inevitably possess, or develop, unstable modes.  The unbounded
growth of instabilities, until second order terms become important,
provides the only viable mechanism to give rise to large-amplitude
spirals no matter how large a number of particles is employed.  With a
recurrence mechanism, such as that described here
(\S\ref{sec.results}), the simulations must support multiple unstable
modes and their superposition naturally leads to the shearing
transient activity that is almost universally reported.  Note that
externally perturbed models, such as those presented by \citet{DVH}
and \citet{KN16}, may not manifest modes if the initial responses
develop before any modes could grow to significant amplitude.

\subsection{Objective of this paper}
\label{sec.aims}
The complicated mechanism just described would be required to account
for the origin of spirals only in models that do not possess any
linear instabilities.  The purpose of this paper is to show how a
recurring cycle of instabilities can arise in models that do possess a
single mild instability at the outset.  We argue that such an
instability cycle is the origin of spiral activity in all simulations,
and hopefully in real galaxies also.

\citet{SC14} also reported that coherent waves appear and decay at
increasing radii and lower frequency over time.  It is possible that
the later instabilities are independent modes caused by uncorrelated
noise fluctuations scratching the DF at larger radii.  However,
\citet{SC14} preferred the idea that the decay of one mode created
conditions to seed a new instability, although they did not provide a
detailed mechanism for recurrence.  Here we describe the recurrence
mechanism in detail.

In fact, \citet{SL89} presented a recurrence mechanism for
self-gravitating spiral modes in low mass particle discs orbiting
around a central mass.  The mechanism they proposed was that
scattering at the OLR created a groove in phase space that excited a
new instability.  We thought it unlikely that the same recurrence
mechanism would work in heavy discs, for reasons that we give in
\S\ref{sec.discuss}, but we here demonstrate that this expectation was
wrong.

\section{Technique}
\label{sec.methods}
Since the dynamical behaviour of fully self-consistent simulations can
be very complicated, we find it fruitful to run simplified simulations
that can capture the phenomena we wish to study without them being
obscured by unrelated activity.  Once understood, it will naturally be
important to show that the behaviour persists under more general
conditions.

Accordingly, we adopt the razor-thin Mestel disc used in the studies
by \citet{Za76} and \citet{To81}, which is characterized by a constant
circular speed $V_0$ at all radii.  Note that since $\Omega_c = V_0/R$
and $\kappa = \surd2V_0/R$, $\Omega_c - \kappa/m$ increases
indefinitely towards the centre except for $m=1$, when it is negative
at all radii.  Thus an ILR will intervene to damp all disturbances
having $\Omega_p>0$ when $m\geq2$, prohibiting possible cavity modes
in a smooth disc, except for $m=1$.

The axisymmetric surface density $\Sigma_0(R) = V_0^2 / (2\pi G R)$
would self-consistently yield the appropriate central attraction for
centrifugal balance.  If the surface density is reduced to fraction
$x$ ($0 < x \leq 1$), with the removed mass added to a rigid halo
to maintain centrifugal balance, then eq.~(\ref{eq.prefm}) implies
that the most vigorously amplified disturbances ($X\simeq1.5$) will
have $m=1.33/x$.  Since \citet{Za76} had found that the full-mass disc
was prone to $m=1$ cavity modes, \citet{To81} preferred a model with
$x=0.5$ in order to avoid lop-sided modes.  Note that although
swing-amplification will be near its most vigorous for $m=2$
disturbances for this choice of $x$, bisymmetric cavity modes in a
smooth disc are disallowed because an ILR would block the feedback
loop.

Toomre employed the DF \citep{To77, BT08}
\begin{equation}
f(E,L_z) = \cases{ x F L_z^q e^{-E/\sigma_R^2} & $L_z>0$ \cr 0 & otherwise, \cr}
\label{eq.zangdf}
\end{equation}
where $q = V_0^2/\sigma_R^2 - 1$ and the normalization constant is
\begin{equation}
F = {1 \over G R_0(R_0V_0)^q} { (q/2 + 0.5)^{q/2+1} \over \pi^{3/2}(q/2-0.5)!}.
\end{equation}
[\citet{Se12} erroneously omitted the factor $R_0^{-(q+1)}$.]
Choosing $q=11.44$ yields a Gaussian distribution of velocities such
that the $x=0.5$ disc has $Q=1.5$.  Toomre further multiplied the DF
$f$ by the double taper function
\begin{equation}
T(L_z) = 
\left[ 1 + \left( {R_0V_0 \over L_z} \right)^\nu \right]^{-1}
\left[ 1 + \left( {L_z \over R_1V_0} \right)^\mu \right]^{-1},
\label{eq.tapers}
\end{equation}
to create a central cut out centered at $R_0$ and an outer taper
centered at $R_1$, while maintaining the centripetal acceleration
$-V_0^2/R$ everywhere.  Setting the taper indices $\nu=4$ and $\mu=5$
yielded an idealized, smooth disc model that Toomre claimed possessed
no small amplitude unstable modes.  We choose $R_1 = 11.5R_0$, and
limit the radial extent of the disc by an energy cut-off that
eliminates particles having sufficient energy to pass $R=20R_0$.  Here
we adopt units such that $V_0 = R_0 = G = 1$.

The evolution of one of the simulations of this model reported by
\citet{Se12} was illustrated in Fig.~\ref{fig.nogroove} above.  The
simulations in the present paper use the same code and disc model,
have the same number of particles (see Table~\ref{tab.params}), but
are integrated to $t = 1000$ only and so are unaffected by the later
rapid growth phase.

\begin{table}
\caption{Numerical parameters}
\label{tab.params}
\begin{tabular}{@{}ll}
Grid size & 106 $\times$ 128 \\
Active sectoral harmonic & 2 or 3 \\
$R_0$ & 8 grid units \\
Softening length & $R_0/8$ \\
Number of particles & $5\times 10^7$ \\
Basic time-step & $R_0/(40V_0)$ \\
Time step zones & 5 \\
Guard zones & 4 \\
\end{tabular}
\end{table}

The particles in our simulations are constrained to move in a plane
over a 2D polar mesh at which the accelerations are calculated, and
interpolated to the position of each particle.  In most cases,
disturbance forces are restricted to a single sectoral harmonic, $m=2$
or $m=3$.  Since the gravitational field is a convolution of the mass
density with a Green function that is most efficiently computed by
Fourier transforms, it is easy to restrict the sectoral hamonics that
contribute to the field when using a polar grid.  A full description
of our numerical procedures is given in the on-line manual
\citep{Se14b} and the code itself is available for download.

Table~\ref{tab.params} gives the values of the numerical parameters
for most simulations reported here, and we note when they are varied
in a few cases.  \citet{Se12} reported, and we have reconfirmed in
this study, that all our results are insensitive to reasonable changes
to grid resoution, time step and zones, and number of particles.
Changes to the softening length do affect the frequencies of the
modes, but not the qualititative behaviour.

To set up each model we draw particles from the tapered DF
(eqs.~\ref{eq.zangdf} and \ref{eq.tapers}) as described in the
appendix of \citet{DS00}.  However, in some experiments
we modified the distribution of selected particles in order to
introduce a single additional feature into the DF, as set out in
\S\ref{sec.groove}.

We also restarted some simulations after ``scrambling'' a copy of the
particle distribution.  By this we mean that we changed
$(R,\phi,v_R,v_\phi) \rightarrow (R,\phi^\prime,v_R,v_\phi)$ for every
particle, with $\phi^\prime$ being chosen at random from a
distribution that is uniform in 0 to $2\pi$.  Clearly scrambling
resets the amplitudes of all non-axisymmetric disturbances back to the
shot noise level of the initial disc, while preserving the radius and
both velocity components (in polar coordinates), so that any features
in the action distribution of the particles that had been introduced
during prior evolution would be preserved.  This is therefore not
equivalent to a fresh start with a different random seed.

Note that the radial action, $J_R \equiv \oint \dot R dR/(2\pi)$
\citep{BT08}, has dimensions of angular momentum, is zero for a
circular orbit, and increases with the eccentricity of the orbit.  In
an axisymmetric potential, the angular momentum, $L_z$, is the other
action for orbits confined to a plane.

\section{Results}
\label{sec.results}
In this section we study the effect of introducing two distinct
features into the otherwise unperturbed disc, both designed to excite
an initial mode of a specific type, namely: a groove mode \citep{SL89,
  SK91} and an outer edge mode \citep{To81, PL89}.  In both cases, we
follow its evolution and study the further modes that are excited.  We
show for each case that the second mode is a true instability of the
modified disc by erasing all non-axisymmetric structure at the time
the second mode began to grow, as just described, and demonstrating
that the scrambled particle distribution possessed a very similar
instability to that which developed in the continued original run.
Furthermore we identify the scattering feature in action space that
was responsible for the second instability, as reported below in
\S\ref{sec.splicing}.

\begin{figure}
\includegraphics[width=.9\hsize,angle=0]{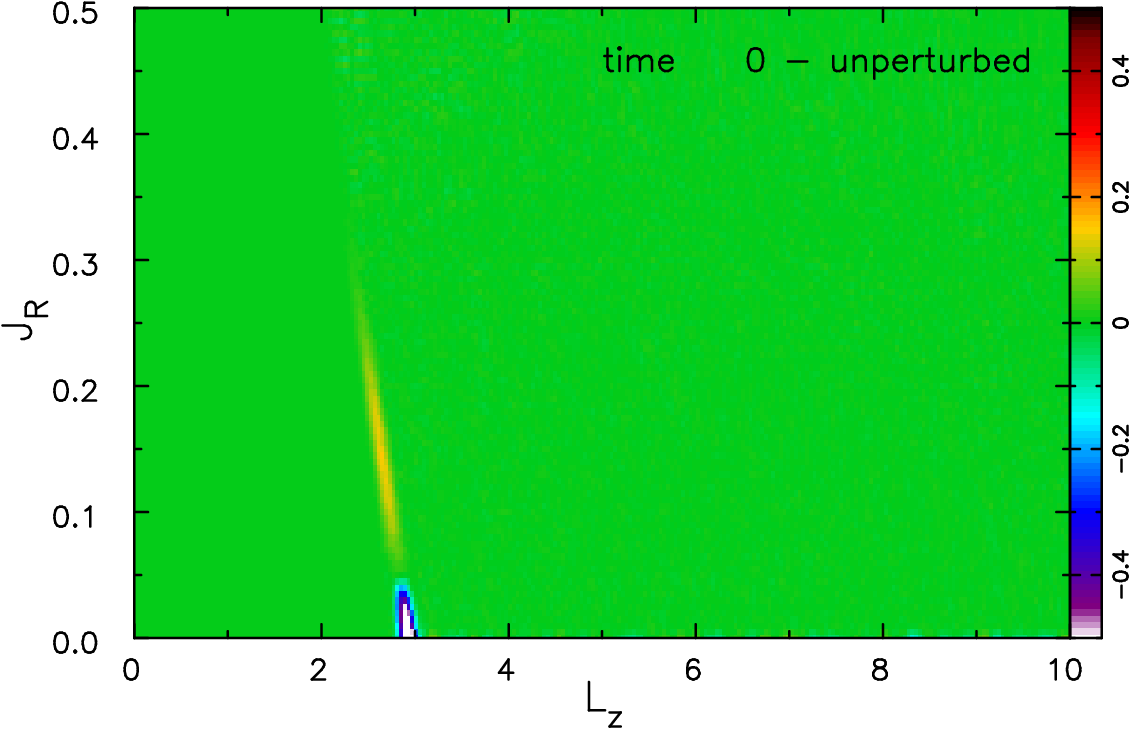}
\caption{The change in the densities of particles in action space
  created by hand in order to insert a groove-like feature into the
  initial model.}
\label{fig.act4864.0-1}
\end{figure}

\subsection{Initial groove mode}
\label{sec.groove} 
We create a deficiency of low-$J_R$ particles over a narrow inclined
range in $(L_z,J_R)$ space by giving some particles a larger $J_R$, as
shown in Fig.~\ref{fig.act4864.0-1}.  The slope of this feature,
$-1/2$, approximately traces the locus of the ILR of a bisymmetric
disturbance having a pattern speed $\Omega_n = 0.1$, although the
disturbance was, in fact, purely hypothetical.  The ILR of a disturbance
having this $\Omega_n$ would lie at $L_z = 2.93$ for circular orbits,
for which $J_R=0$.\footnote{Note that all the major resonances with
  any pattern speed and any angular periodicity in the self-similar
  Mestel disc lie on lines of slope $\simeq -1/2$ in the space of
  these actions, for $J_R \ll L_z$.}

We describe this feature as a groove, since it is a deficiency at low
$J_R$ over a narrow range in $L_z$.  Since orbits in this self-similar
disc have typical epicyclic radii of $\sigma_R/\kappa \simeq 0.2R_g$
about the guiding centre radius, this narrow groove in action space
causes no noticeable change to the surface density profile.

\begin{figure}
\includegraphics[width=.9\hsize,angle=0]{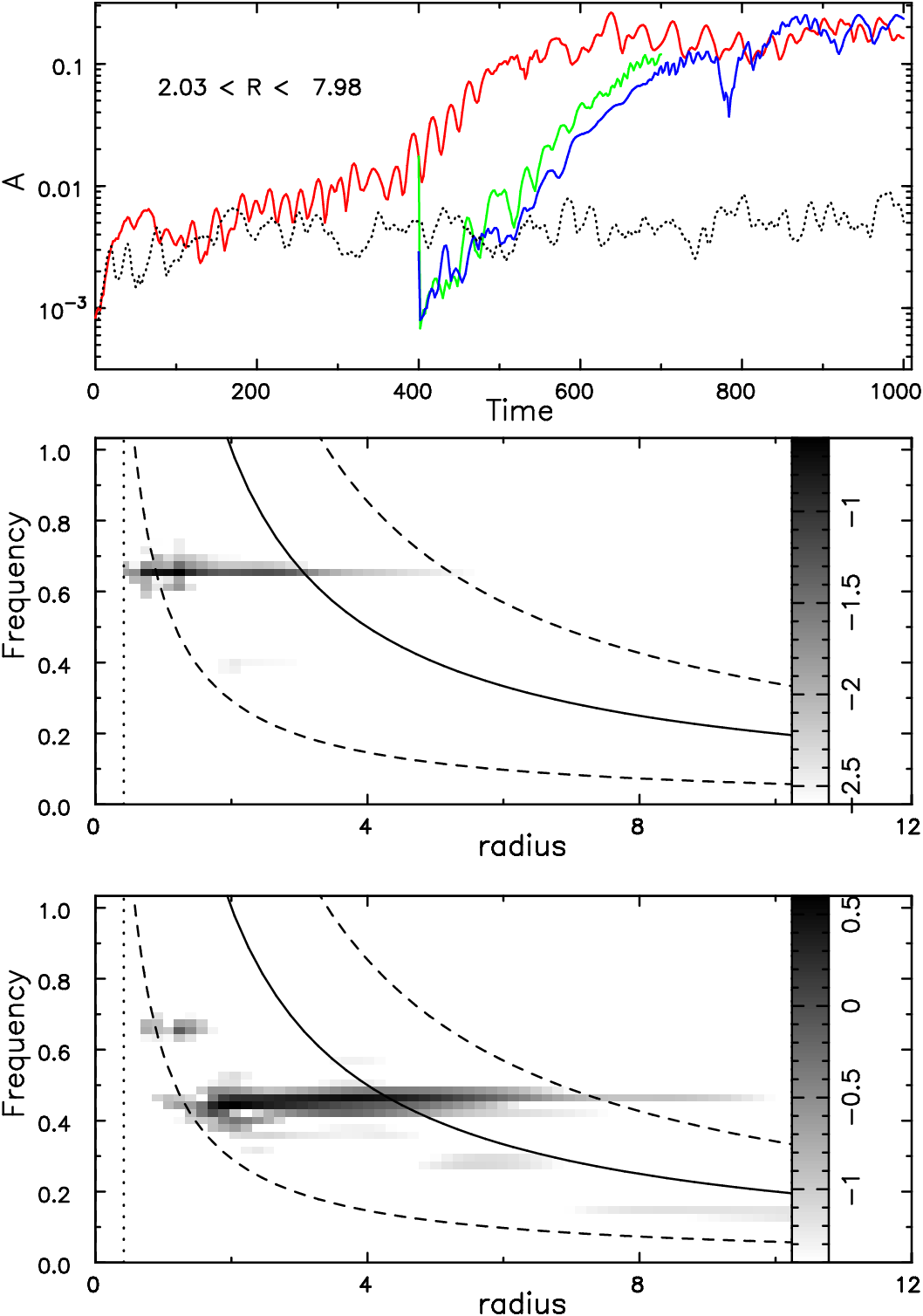}
\caption{The evolution of model G, which was seeded with the groove
  shown in Fig.~\ref{fig.act4864.0-1}.  The red line in the top panel
  shows the amplitude evolution, the dotted black line indicates what
  would have happened without the groove, and explanations of the
  green and blue lines given in \S\S\ref{sec.g2} \& \ref{sec.splicing}
  respectively.  The rapid, but small, variations in the amplitude
  reflect alternating constructive and destructive interference by
  uncorrelated episodes of swing-amplified noise, that have longer
  periods at later times as this behaviour spreads to the outer part
  of the radial range.  The middle and bottom panels, for which again
  the grey scales differ, present power spectra over the time ranges
  $100\leq t \leq 400$ and $400\leq t \leq 700$ respectively, which
  each manifest a single mode.}
\label{fig.grevol}
\end{figure}

\begin{table*}
\begin{tabular}{@{}rllcccccc}
Line \# & Model & $m$ & Mode \# & time range & $\omega = m\Omega_p+i\beta$ & $R_{\rm ILR}$ & $R_{\rm CR}$ & $R_{\rm OLR}$ \\
1 & G & 2 & 1 & $100 \leq t \leq 200$ & $0.656\pm0.005 + (0.017\pm0.003)i$ & $\sim 0.89$  & $\sim 3.05$  & $\sim 5.20$ \\
2 & G & 2 & 2 & $200 \leq t \leq 500$ & $0.393\pm0.003 + (0.022\pm0.002)i$ &  & $\sim 5.09$  & \\
3 & GR & 2 & 1 & $400 \leq t \leq 700$ & $0.398\pm0.001 + (0.019\pm0.002)i$ &  & $\sim 5.03$  & \\
4 & GR & 2 & 2 & $400 \leq t \leq 700$ & $0.672\pm0.001 + (0.013\pm0.004)i$ &  & $\sim 2.98$  & \\
5 & GS$_{\rm OLR}$ & 2 & 1 & $400 \leq t \leq 700$ & $0.400\pm0.008 + (0.019\pm0.005)i$ &  & $\sim 5.00$  & \\
6 & GS$_{\rm CR}$ & 2 & 1 & $400 \leq t \leq 700$ & $0.662\pm0.030 + (0.010\pm0.008)i$ &  & $\sim 3.02$  & \\
7 & E & 3 & 1 & $300 \leq t \leq 600$ & $0.233\pm0.001 + (0.005\pm0.001)i$ & $\sim 6.8$ & $\sim 12.8$  & $\sim 18.9$ \\
8 & E & 3 & 2 & $500 \leq t \leq 780$ & $0.417\pm0.001 + (0.020\pm0.006)i$ &  & $\sim 7.2$  & \\
9 & ER & 3 & 1 & $600 \leq t \leq 800$ & $0.423\pm0.001 + (0.014\pm0.001)i$ &  & $\sim 7.1$  & \\
10 & ER2 & 2 & 1 & $600 \leq t \leq 900$ & $0.271\pm0.002 + (0.013\pm0.002)i$ &  & $\sim 7.4$  & \\
11 & ER2 & 2 & 2 & $600 \leq t \leq 900$ & $0.136\pm0.006 + (0.022\pm0.004)i$ &  & $\sim 14.7$  & \\
12 & GS3$_{\rm OLR}$ & 3 & 1 & $400 \leq t \leq 680$ & $0.614\pm0.009 + (0.013\pm0.001)i$ &  & $\sim 4.9$  & \\
\end{tabular}
\caption{Summary of the modes fitted to data from simulations
  described in \S\ref{sec.results}, including the fitted complex
  frequencies, $\omega$ with $\beta$ being the growth rate of the
  mode, and the radii of the principal resonances for circular orbits
  given by $(m+l\surd2)/\Re(\omega)$, although we leave uninteresting
  entries blank for clarity.}
\label{tab.modes}
\end{table*}

Our, perhaps somewhat clumsy, procedure for shifting particles was
motivated by our observations of resonance scattering in previous
simulations.  We calculated the frequency distance from the resonance
$\delta\omega = 2\Omega_\phi - \Omega_R - 2\Omega_n $ for each
particle, where $\Omega_\phi$ and $\Omega_R$ are respectively the
angular frequencies of the guiding centre and radial oscillation for
orbits of arbitrary eccentricity; note that these frequencies tend to
the familiar quantities $\Omega_\phi \rightarrow \Omega_c$ and
$\Omega_R \rightarrow \kappa$ as the eccentricity $\rightarrow 0$.  We
then select particles that have $g^2 = 1 - (\delta\omega / w_\omega
\Omega_n)^2 > 0$ as candidates to be scattered.  The relative
frequency width, $w_\omega = 0.06$, of this feature is deliberately
rather narrow.  Of these, we further select only those particles whose
energy of random motion $E_r = E - E_c(L_z) < g E_{r,{\rm lim}}$,
where the groove extends to the limiting random energy $E_{r,{\rm
    lim}} = 0.025$ at $g=1$, \ie\ along the resonance locus.  The
probability that these selected particles are given larger random
energy decreases from $d$ to zero as $E_r$ increases from zero to this
maximum, where the fractional depth of the groove $d = 0.25$.  The new
$E_r = E_{r,{\rm lim}}(1+3p)$ where $p$ is the solution of
$\exp(-p^2)=s$, with $s$ being randomly drawn from a distribution
uniform in 0--1, and finally $\Delta L_z = \Delta E / \Omega_n$.

A simulation that employed this modified DF, model G, supported a
groove instability at first.  The red curve in the top panel of
Fig.~\ref{fig.grevol} presents the time evolution of the ratio
$\langle\Sigma_2(R)/\Sigma_0(R)\rangle$ averaged over the radius range
$2<R<8$.  Since this is a different measure from that presented in
Fig.~\ref{fig.nogroove}, the dotted curve indicates this measure for
the same model without the groove.  The green and blue lines are
explained below.  Power spectra over two periods of growth are shown
in the lower two panels.  The middle panel is for the same time period
as the middle panel in Fig.~\ref{fig.nogroove}, although the grey
scales differ, and the contrast in the appearance is quite striking.

\subsubsection{First mode}
\label{sec.g1}
The disturbance in model G grew slowly until $t\sim400$, and the power
spectrum over this time interval in the middle panel of
Fig.~\ref{fig.grevol} is dominated by a coherent disturbance of
angular frequency $2\Omega_p \approx 0.65$ that is mostly localized
between the ILR and corotation.  A couple of other frequencies that
are probably due to swing-amplified noise are also faintly visible,
and these uncorellated mild disturbances are responsible for the small,
but rapid, amplitude fluctuations in the red line in the top panel.

\begin{figure}
\includegraphics[width=.9\hsize,angle=0]{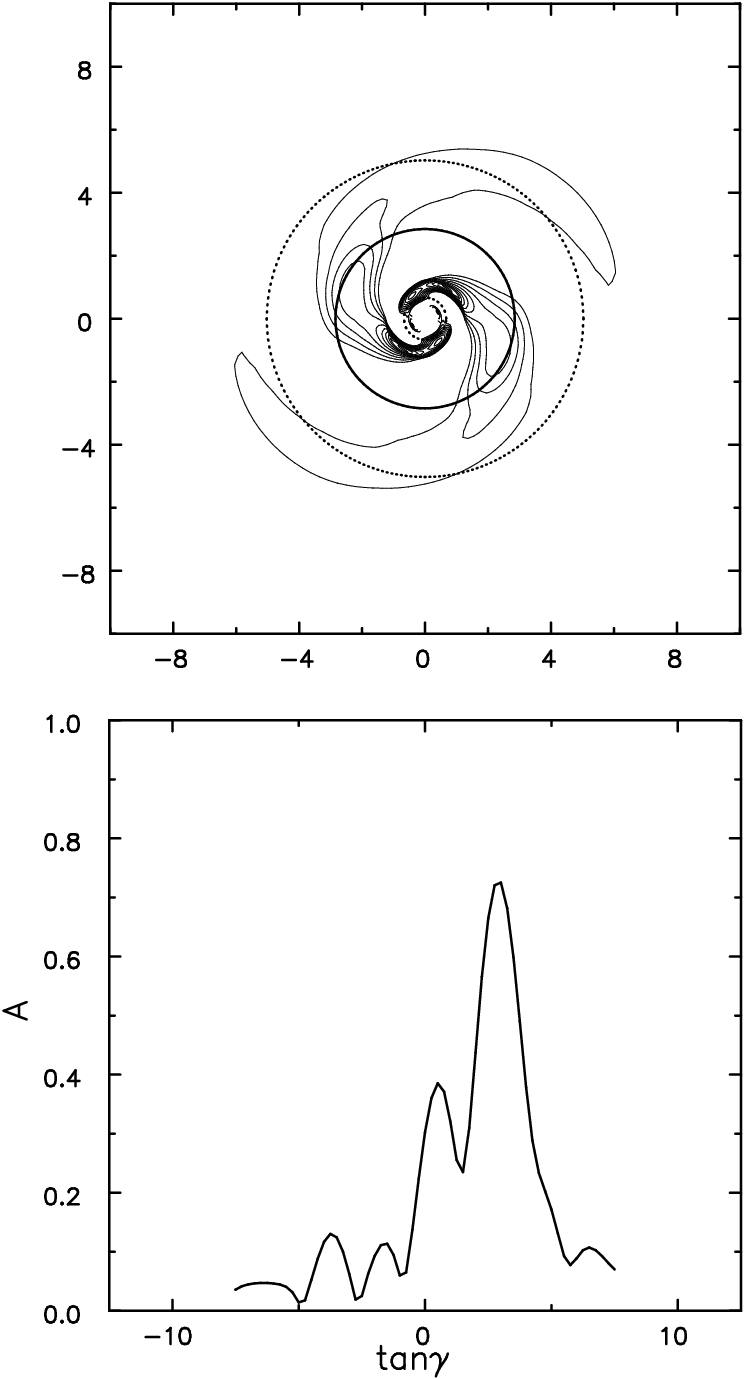}
\caption{Top panel: contours of the relative over density of the mode
  fitted to the results from model G; distances are marked in units of
  $R_0$ and the circles mark the radii of the principal resonances for
  circular orbits, given in line 1 of Table~\ref{tab.modes}.  Bottom
  panel: the logarithmic spiral transform of the mode density and as
  in Fig.~\ref{fig.mode4169}, the amplitude scale is arbitrary.}
\label{fig.mode4864}
\end{figure}

We employed the mode-fitting procedure devised by \citet{SA86} to the
data from model G over the time interval $100 \leq t \leq 400$ and
give its estimated eigenfrequency in line 1 of Table~\ref{tab.modes}.
The shape of the fitted mode, with its principal resonances for
circular orbits ($J_R=0$) marked by the circles, is shown in the upper
panel of Fig~\ref{fig.mode4864}, and the decomposition of the mode
into logarithmic spirals is in the lower panel.

The mechanism for the groove instability was fully explained by
\citet{SK91} and outlined above in \S1.1.  It is not a cavity-type
mode, and the frequency is determined by the gradients in angular
momentum density for orbits of small radial action.  The density
changes caused by non-axisymmetric disturbances at the groove
``edges'' excite a supporting response from the surrounding disc,
making it a large-scale spiral mode.  In principle, the circular
angular frequency of the groove centre should be the pattern speed of
the mode, but geometric factors in low-$m$ modes and finite growth
rate cause the corotation radius to lie slightly farther out.  The
shape of the mode is determined by the shear rate and surface density
profile of the disc, so that in the present self-similar disc, groove
modes should have closely similar shapes and pitch angles and their
spatial scale will vary with the corotation radius.

Evidence that this mode is a groove instability \citep{SL89} is
two-fold.  First, corotation (line 1 of Table~\ref{tab.modes}) lies close
to, but just outside \citep{SK91}, the groove centre at $L_z = 2.93$.
Second, the leading component in the mode transform (bottom panel of
Fig.~\ref{fig.mode4864}) is much weaker than that in
Fig.~\ref{fig.mode4169} because groove modes (\S\ref{sec.modes}) do
not operate by a feedback cycle of combined leading and trailing
waves.

Our choice of a very narrow frequency width for the groove ensured a
low growth rate for the mode \citep{SK91}.  We found in other
simulations, not described in detail here, that wider grooves excited
a mode that grew more vigorously to higher amplitude.  We preferred to
study an instability that caused mild non-linear changes in order to
make it easier to follow further activity that was a consequence of
the initial mode.

\subsubsection{Second mode}
\label{sec.g2}
After the slow growth to $t=400$, the disturbance amplitude in model
G, shown by the red line in the top panel of Fig.~\ref{fig.grevol},
rose more rapidly over the time range $400 < t < 700$.  The power
spectrum over this time range in the bottom panel revealed a new
coherent wave having a frequency $\sim 0.45$.  As the mode had clearly
saturated by the end of this period, we fitted a mode over the time
interval $200 < t < 500$, obtaining good fits with eigenfrequencies in
the range given in line 2 of Table~\ref{tab.modes}.

In order to verify that this second mode was a true linear instability
of the modified disc, we made a copy of the particle distribution at
$t=400$ and evolved the scrambled (see \S\ref{sec.methods}) particle
distribution in a separate simulation, designated GR.  The amplitude
evolution in the simulation starting from this scrambled disc is shown
by the green line in the top panel of Fig.~\ref{fig.grevol}.  The very
first value was measured from the particles before they were
scrambled, after which the amplitude rose out of the shot noise
steadily to $t=700$ when the run was stopped.

We found that the evolution of the scrambled disc, model GR, was again
dominated by an exponentially growing disturbance of fixed pattern
speed.  The eigenfrequency fitted to the data from this run, given in
line 3 of Table~\ref{tab.modes}, is within the uncertainties the same
as that in line 2 -- the second mode found in model G.  A second,
milder instability was also present that had the fitted frequency
given in line 4.  At first we thought it could be the original groove
mode, although the pattern speed is distinctly higher, but we identify
its progeny at the end of \S\ref{sec.splicing}.

\begin{figure}
\includegraphics[width=.9\hsize,angle=0]{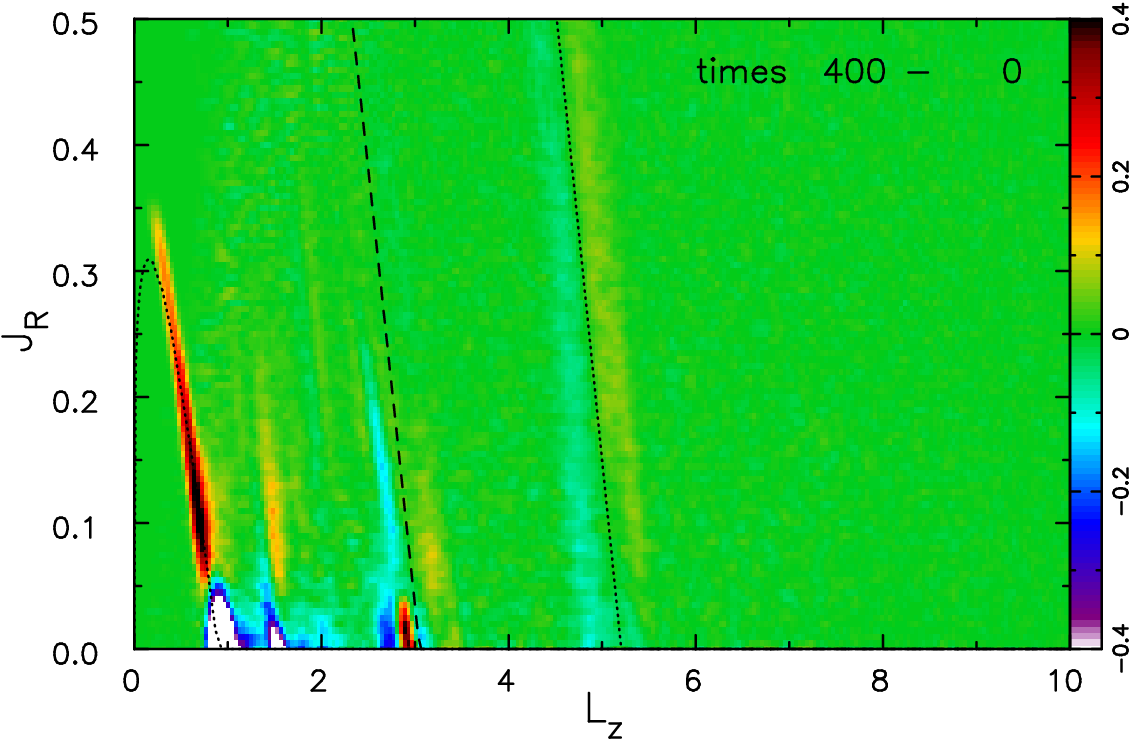}
\caption{Difference between the densities of particles in action space
  from $t=0$ to $t=400$ in model G, started with an initial groove.
  The dashed line shows the locus of the corotation resonance for the
  measured mode frequency, and the dotted lines are the Lindblad
  resonances.  The calculation of actions and the loci of the
  resonances in the evolved model here, and in all subsequent cases,
  neglected any non-axisymmetry in the potential.}
\label{fig.act4864.400-0}

\medskip
\includegraphics[width=.9\hsize,angle=0]{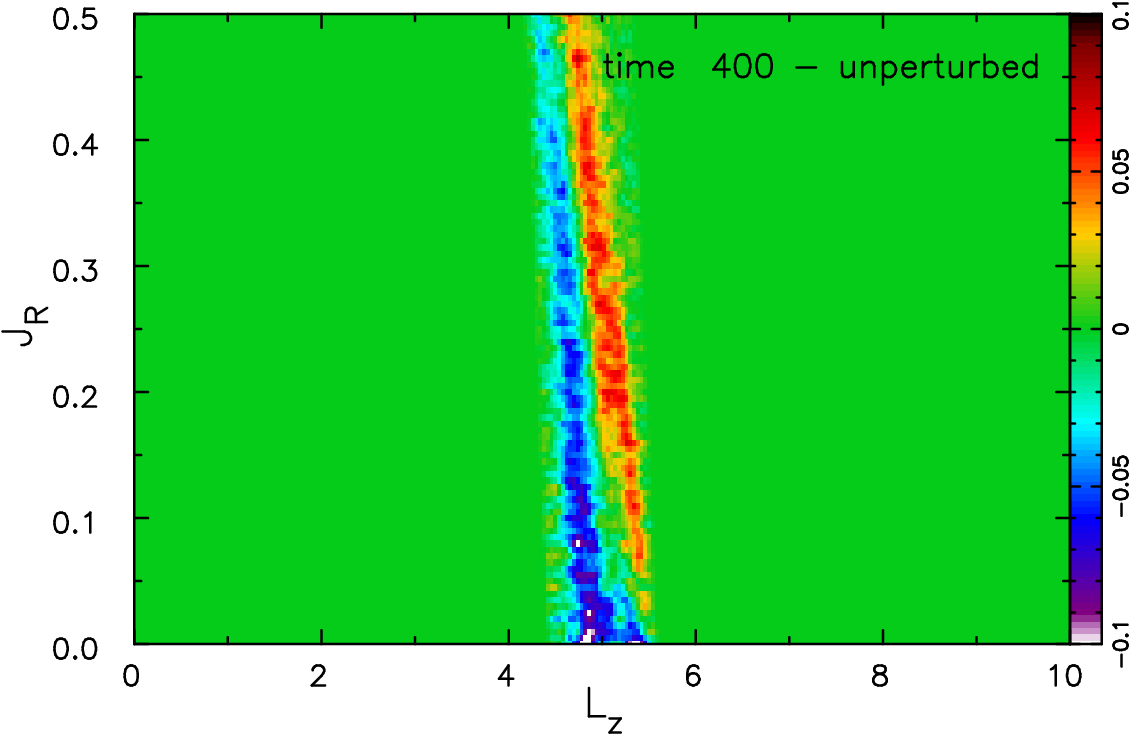}
\caption{Difference from a pristine undisturbed DF and that created by
  splicing those particles in the OLR scattering feature from
  Fig.~\ref{fig.act4864.400-0}.  Note that the colour contrast has
  been enhanced in order to show how nearly seamless the splicing
  process was.  These particles were employed in model GS$_{\rm OLR}$.}
\label{fig.act4871.400-1}
\end{figure}

\subsubsection{Cause of the second mode}
\label{sec.splicing}
The second more vigorous mode, which was not present in the early
evolution of the same model G (middle panel of Fig.~\ref{fig.grevol}),
was caused by changes to the DF produced by the first mode.
Fig.~\ref{fig.act4864.400-0} reveals the changes to the density of
particles in action space between times 0 and 400.  The colour scale
in this figure shows where the density of particles at $t=400$ is both
greater and less than that at $t=0$, and reveals several features.
The dashed line marks the locus of the corotation resonance, and the
dotted lines the Lindblad resonances, for the fitted frequency of the
first mode in this simulation.  These lines indicate that the most
prominent feature is the result of ILR scattering and the faint pair
of features at larger $L_z$ are associated with the OLR.  A paired
deficiency and excess have appeared near the location of the original
groove at $L_z \simeq 2.93$, reflecting the changes that must have
been caused by the instability.  As such changes occur near
corotation, marked by the dashed line, they should not cause any
significant change to $J_R$ \citep{SB02}, which is largely true.
Since the groove instability is so mild, Fig.~\ref{fig.act4864.400-0}
also shows the effects of scattering by weak noise features that
generally had lower frequencies than the excited mode.

Note the differences in the nature of the changes caused by scattering
at the ILR from those at the OLR.  In action space, the scattering
vector at a resonance has slope $\Delta J_R/\Delta Lz = l/m$
\citep{SB02}, where $m$ is the angular multiplicity of the pattern and
$l=0,\;\mp1$ at respectively corotation and the inner and outer
Lindblad resonances.  Formally the slope $l/m$ is exact only as $J_R
\rightarrow 0$ but we find it is an excellent approximation over the
range of $J_R$ of interest here.  As noted before \citep{Se12, SC14,
  Se19}, the scattering vector slope, $-1/2$, is almost perfectly
aligned with the ILR locus for $m=2$ waves, implying that particles
stay on resonance as they gain $J_R$, allowing large changes to build
up.  Thus ILR scattering moves particles along the resonance locus,
creating a deficiency of particles at low $J_R$, and an excess for
larger $J_R$, just as we created artificially at the start of the
first simulation (Fig.~\ref{fig.act4864.0-1}).  On the other hand,
scattering vectors at the OLR have positive slope $+1/m$, while the
resonance locus again has slope $-1/2$.  Thus we see particles are
scattered across the resonance from a sloping deficiency to an almost
parallel line where there is a slight excess.  It is perhaps
interesting that the lines are approximately parallel, indicating that
the action changes are roughly independent of the initial $J_R$.  More
importantly, $J_R$ is an intrinsically positive quantity, and
therefore Lindblad resonance scattering always creates a deficiency at
$J_R \ga 0$.

\begin{figure*}
\includegraphics[width=.9\hsize,angle=0]{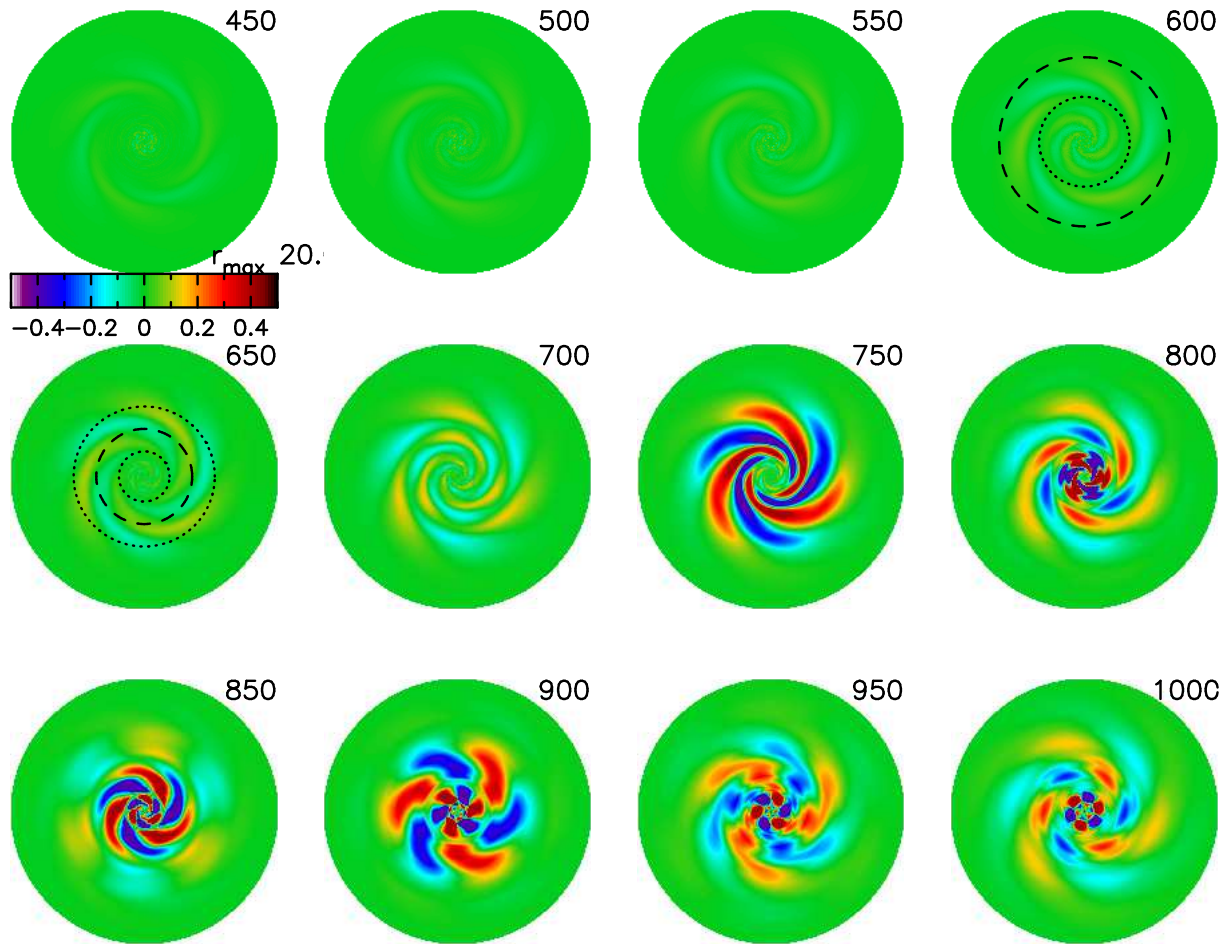}
\caption{The later part of the evolution of model E having a steeper
  outer edge.  The orbit period at the corotation radius of the edge
  mode ($R=12.8$) is $\sim80$ in these units.  The circles at $t=600$
  mark corotation and the ILR of the edge mode, those at $t=650$ are
  for corotation and both Lindblad resonances for the second mode.
  The indicated colour scale shows the absolute over- or under-density
  of non-axisymmetric features, which has been scaled up by a factor
  of 100.}
\label{fig.cntd4887}
\end{figure*}

We suspected that one of the features in Fig.~\ref{fig.act4864.400-0}
was responsible for the new instability in model G.  In order to
determine which, we extracted all the particles in one of the features
and spliced them into a pristine undisturbed particle distribution
that we then evolved.  In detail, we began the splicing procedure by
creating a fresh set of particles from the doubly tapered Toomre-Zang
DF described in \S\ref{sec.methods} that lacked any additional
features; recall that \citet{To81} had predicted, and \citet{Se12} had
confirmed, that this model has no coherent, small amplitude
instabilties.  We then extracted all the particles from the evolved
distribution at $t=400$ that lay within a trapezium in action space of
width $\Delta L_z = 1$ that was defined by $|L_z + 0.5 J_R - L_{z,0}|
< 0.5$, for some value of $L_{z,0}$.  We spliced the extracted
particles into the undisturbed set by substituting the values of
$(x,y,v_x,v_y)$ of these selected particles from the evolved set for
the values in the undisturbed set of particles that lay within the
same trapezium in action space.\footnote{Since the central attraction,
  $-V_0^2/R$, is an unchanging function throughout these experiments,
  we do not have to worry about possible changes to the axisymmetric
  potential.} This process required attention to two points in order
to obtain a near seamless splice.  First, almost all the particles in
the evolved model had changed their $L_z$ values during the evolution
to $t=400$, making it impossible to just replace particle $n$ from the
pristine set with particle $n$ from the evolved set, even though they
were both originally drawn from the same DF by the same algorithm.  We
therefore had to compute $J^\prime = L_z + 0.5J_R$ for every particle
in both sets and rank order them to identify those in the desired
range.  Second, the large numbers of particles in the desired range of
$J^\prime$, typically some 4.5 million, was not precisely the same in
the two sets, so that we had to skip the occasional particle, usually
one in several hundred, in the larger set to ensure equal numbers that
spanned the entire range.  Fig.~\ref{fig.act4871.400-1} shows the
difference between the spliced and undisturbed sets for $L_{z,0}=5$.
It is clear that the OLR scattering feature from
Fig.~\ref{fig.act4864.400-0} has been satisfactorily isolated from all
other features in action space at that time, without introducing sharp
edges to the splice.

The time evolution of the disturbance density in a simulation started
from the spliced distribution, model GS$_{\rm OLR}$, is indicated by
the blue line in the top panel of Fig.~\ref{fig.grevol}.  The mild
non-axisymmetry of the spliced particles at $t=400$, was erased by
scrambling before the evolution commenced, and the disturbance density
grew out of the noise, closely following the evolution from the simply
scrambled disc, model GR, traced by the green line.\footnote{Other
  modes are excited after $t\sim700$, and interference between
  disturbances rotating at different rates can cause large temporary
  decreases in the measured amplitude.} The eigenfrequency fitted to
the data from this spliced run is given in line 5 of
Table~\ref{tab.modes} and is, again within the uncertainties, the same
as those in lines 2 and 3, the second mode in model G and the mode in
the scrambled disc, model GR.  This evidence conclusively shows that
the OLR scattering feature of the first mode in model G was
responsible for creating the second instability.

Note that the corotation resonance for near circular orbits of the
modes given in lines 2, 3, and 5 of Table~\ref{tab.modes}, all lie
near the OLR feature created by the first mode
(Fig.~\ref{fig.act4871.400-1}) strongly suggesting that the second
mode was also a groove instability.  It seems likely that the deficit
of particles near $J_R \ga 0$ was the cause.

We were quite surprised to find it was the OLR scattering feature that
was responsible for the second instability for the reasons set out in
the discussion below (\S\ref{sec.discuss}).  Yet our experiment with
the isolated weak scattering feature of the OLR leaves no doubt that
it did indeed provoke a new instability that was more vigorous than
the original mode.

The near coincidence of the OLR for the first mode (line 1 of
Table~\ref{tab.modes}) with CR for the second (lines 2, 3 and 5 of
Table~\ref{tab.modes}) suggests non-linear mode coupling.  Resonance
scattering to create the new groove is a non-linear effect, but we
have strong evidence against the hypothesis of coupling at resonances
between large amplitude waves that was first proposed by \citet{Ta87}.
A technical reason is that their theory requires the interaction of
three waves, with the third wave having rotational symmetry that is
the sum or difference of the two other waves, and our simulation in
which disturbance forces were restricted to $m=2$ could not support
disturbances having $m=0$ or 4.  However, the more compelling argument
is that scrambling the particles erased all non-axisymmetric features
in the density distribution, which need to have significant amplitude
for the mode coupling mechanism to work.  Instead, our simulations
provide clear evidence that the second mode is a true, linear
instability of the modified disc at $t=400$.

We tried other experiments for which we chose other splice centres,
but do not illustrate the results.  Choosing $L_{z,0}=3$, model
GS$_{\rm CR}$\Ignore{run4870} supported a mild groove mode with
frequency given in line 6 of Table~\ref{tab.modes}; the deficiency of
the original DF had been shifted to lower $L_z$ by the evolution of
the first groove mode, causing the slightly higher pattern speed than
that reported in line 1.  It is satisfying to note that this
instability approximately matches that of the second mode in the
scrambled model GR, reported in line 4 of the table.  Simulations of
particle distributions that resulted from splicing with
$L_{z,0}=1$\Ignore{run4872} or from $L_{z,0}=2$\Ignore{run4873}
appeared not to support any coherent instability.  The strong ILR
scattering feature from the first mode lies within the inner taper,
and is therefore unable to excite a significant instability because
the tapered surounding disc would provide only a weak supporting
response of \citep{SK91}.

\subsection{Initial edge mode}
\label{sec.edge} 
All we needed to do to excite an edge mode was to increase the value
of $\mu$ for the outer index taper (eq.~\ref{eq.tapers}).  Setting
$\mu=20$ created a vigorous instability that saturated at large
amplitude, strongly distorting a large fraction of the disc.  After
some experimentation, we found that setting $\mu=12$, model E, yielded
an outer edge mode with a moderate growth rate that enabled us to
understand the subsequent evolution.  Note that for the sequence of
runs reported in this section, in which the initial disturbances were
in the outer disc where the grid becomes coarser, we used a 4 times
finer grid ($404 \times 512$).  We also restricted disturbance forces
to $m=3$, rather than $m=2$, in order that the modes have smaller
radial extent, but the other numerical parameters were unchanged.

\begin{figure}
\includegraphics[width=.9\hsize,angle=0]{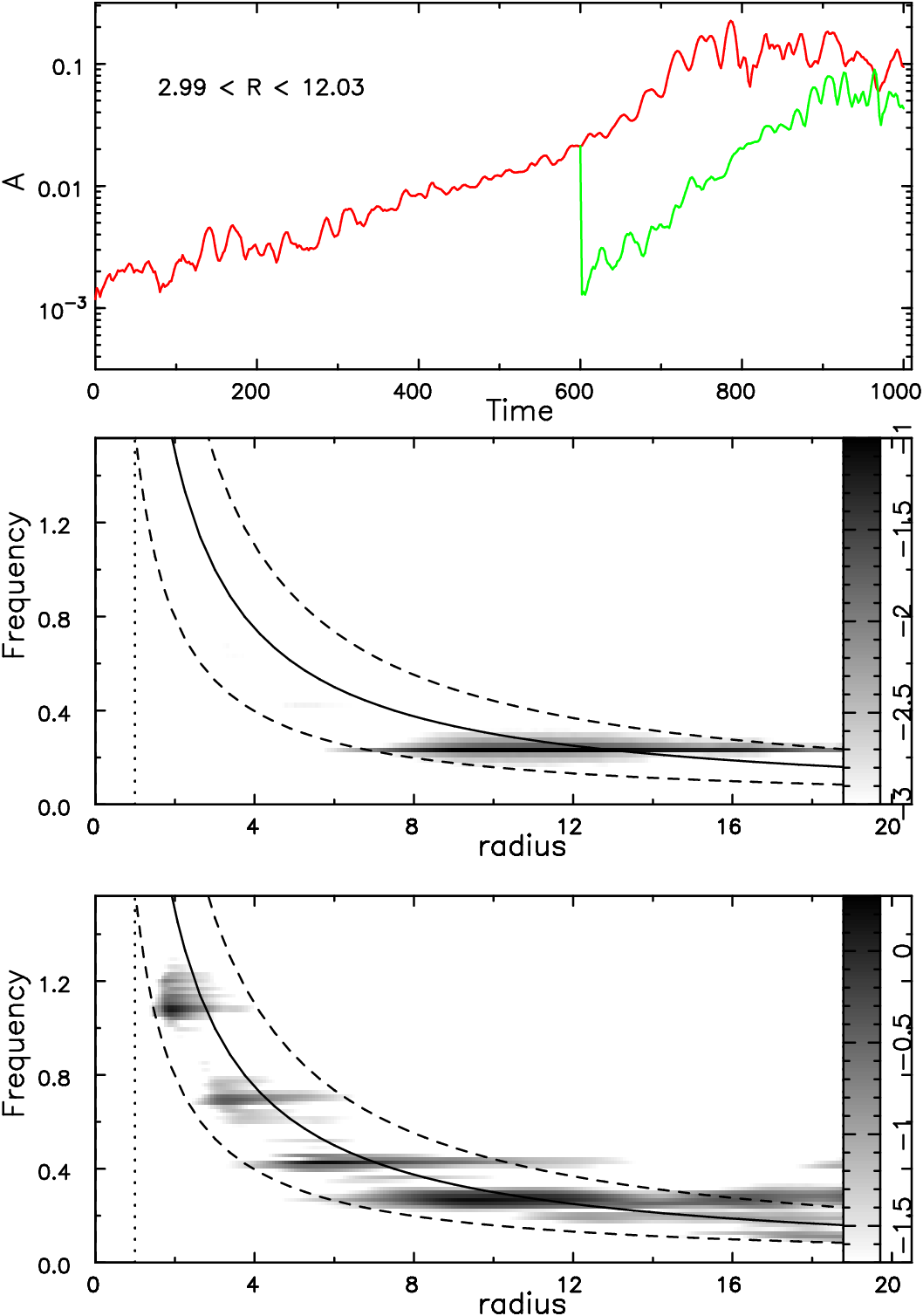}
\caption{The evolution of model E having a steeper outer edge.  The
  red line in the top panel shows the amplitude evolution, while the
  middle and bottom panels show power spectra over the time ranges
  $350\leq t \leq 650$ and $600\leq t \leq 1000$ respectively and
  again the grey scales differ.}
\label{fig.edgevol}
\end{figure}

The evolution of the non-axisymmetric density of model E is shown in
Fig.~\ref{fig.cntd4887}.  A 3-arm spiral grows slowly in the outer
disc until $t\sim600$, after which additional patterns appear closer
to the disc centre.  The amplitude evolution of the $m=3$ disturbance
density over the radius range $3<R<12$ and power spectra over two
periods of growth are shown in Fig.~\ref{fig.edgevol}.

\begin{figure}
\includegraphics[width=.9\hsize,angle=0]{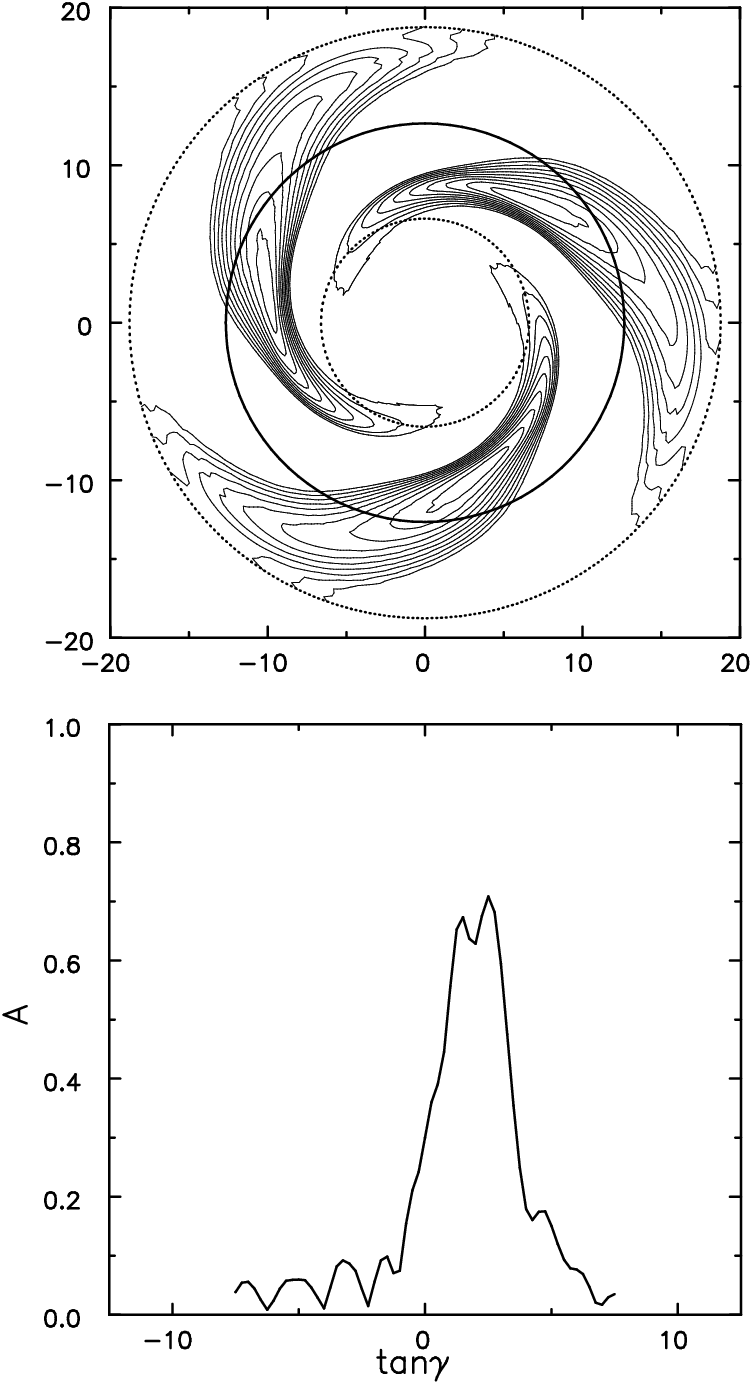}
\caption{Top panel: contours of the relative over density of the mode
  fitted to the simulation results of model E given in line 7 of
  Table~\ref{tab.modes}.  Distances are marked in units of $R_0$ and
  the circles mark the radii of the principal resonances for circular
  orbits.  Bottom panel: the logarithmic spiral transform of the mode
  density.  Again the amplitude scale is arbitrary.}
\label{fig.mode4887}
\end{figure}

\subsubsection{First mode}
\label{sec.e1}
The evolution of model E to $t=600$, is dominated (middle panel of
Fig.~\ref{fig.edgevol}) by a single, exponentially growing disturbance
of frequency $3\Omega_p \ga 0.2$, that extends from its ILR to OLR,
with corotation perhaps near $R \sim 13$.  The mode fitting software
yielded an eigenfrequency given in line 7 of Table~\ref{tab.modes} and
the fitted mode and its logarithmic spiral transform are presented in
Fig.~\ref{fig.mode4887}.

It has the hallmarks of an edge mode.  As expected \citep{To89},
$R_{\rm CR}$ (line 7 of Table~\ref{tab.modes}) lies outside the radius
of the steepest gradient, which is at $L_z=11.5$, the centre of the
outer taper.  Again, the logarithmic spiral transform of the mode,
lower panel of Fig.~\ref{fig.mode4887}, has no significant leading
component as expected for an edge mode, since the mechanism does not
require feedback through leading waves (see \S\ref{sec.modes}).

\begin{figure}
\includegraphics[width=.9\hsize,angle=0]{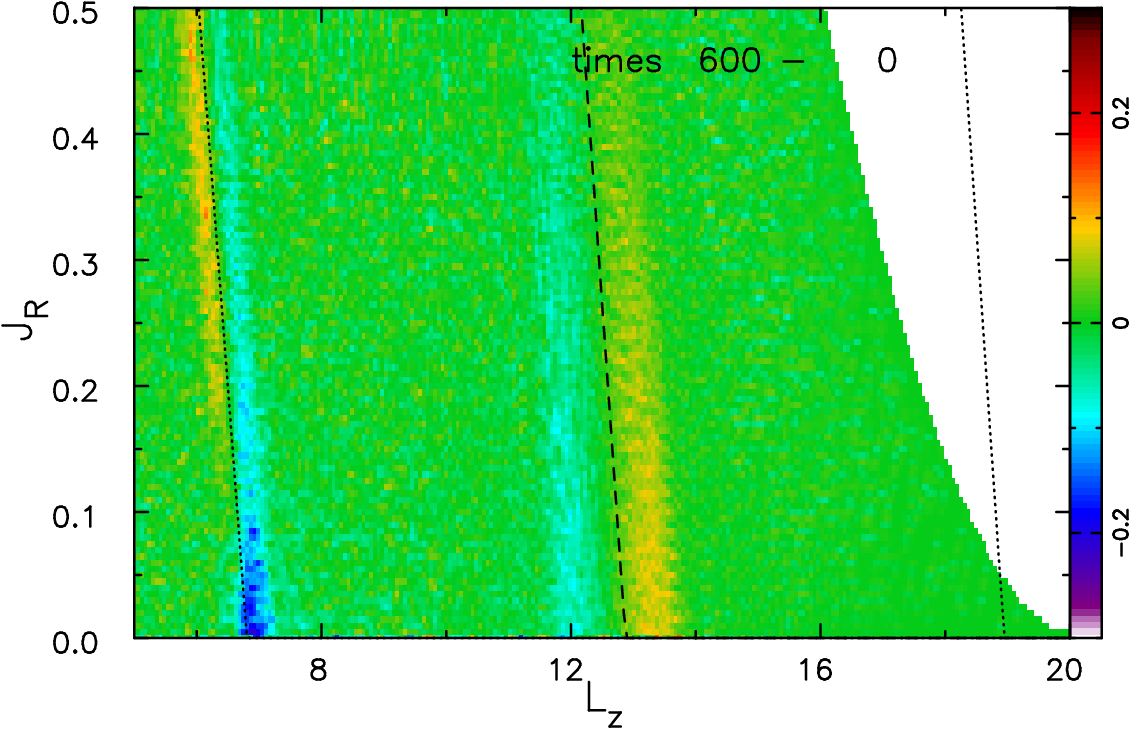}
\caption{Difference between the densities of particles in action space
  from $t=0$ to $t=600$, from model E having the sharper outer edge.
  The dashed line shows the locus of the corotation resonance for the
  measured mode frequency, line 7 of Table~\ref{tab.modes}, and the
  dotted lines are the Lindblad resonances.  Note that the range of
  abscissae differs from other figures of this type and our initial
  upper energy bound (\S\ref{sec.methods}) excluded particles from the
  white region.}
\label{fig.act4887.600-0}
\end{figure}

Fig.~\ref{fig.act4887.600-0} displays the changes to the density of
particles in action space between $t=0$ and $t=600$ in model E, and
the dashed and dotted lines mark the loci of the principal resonances
for the measured pattern speed of the original edge mode.  It reveals
scattering features at corotation and the ILR, but there are no
significant changes at the OLR because there is very little disc left
at those radii, and angular momentum transport by this edge mode is
from the ILR to CR only.  Since the power spectra
(Fig.~\ref{fig.edgevol}) and the fitted mode shape
(Fig.~\ref{fig.mode4887}) were drawn using the relative overdensity at
each radius, they exaggerate the apparent amplitude of the mode in the
very outer part of the disc.  The very low frequency features in the
bottom panel of Fig.~\ref{fig.edgevol} have a small absolute
amplitude, and probably arise from shot noise due to the low particle
density near the disc edge.

As noted above, scattering vectors in action space have slopes $=l/m$.
Thus the near perfect alignment of the scattering vector with the
locus of the ILR occurs only for $m=2$ disturbances, and therefore the
excess and deficiency in Fig.~\ref{fig.act4887.600-0} at the ILR are
slightly offset from the resonance locus for this $m=3$ pattern, and
the groove carved by the mode at $L_z=6.9$ is fractionally exterior to
its ILR.

\subsubsection{An inwardly propagating cascade of groove modes}
\label{sec.e2}
The power spectrum of the later part of this simulation, in the bottom
panel of Fig.~\ref{fig.edgevol}, indicates several additional coherent
waves, in addition to the continuing presence of the first mode, which
had not even finished growing by $t=600$.

We have fitted two modes to the data from the simulation over the
period $500 < t < 780$, recovering both the edge mode and the second
mode, which has an estimated frequency given in line 8 of
Table~\ref{tab.modes}.  Corotation for this second mode, also given in
line 8, is somewhat outside the groove centre carved by the original
edge mode, as is usual groove mode \citep{SK91}.

The green line in Fig.~\ref{fig.edgevol} shows the amplitude evolution
we obtain in a new simulation, model ER, started from the scrambled
particles of model E at $t=600$.  Once again we find the perturbation
amplitude grows rapidly with the estimated frequency given in line 9
of Table~\ref{tab.modes}, which is in reasonable agreement with the
frequency of the second mode given in line 8.  Note the second mode
outgrows the original edge mode, which is also detectable in the
scrambled simulation.  The true uncertainties in the measured
frequencies and resonance radii in this case are probably greater then
the formal values given in the table because we were attempting to fit
two vigorously growing modes over a short time interval.

The numerical evidence just presented implies that the second mode in
model E is a groove mode seeded by ILR scattering by the first mode
(Fig.~\ref{fig.act4887.600-0}).  Once again, the near coincidence of
the ILR of mode 1 (line 7 of Table~\ref{tab.modes}) is CR of its
daughter (line 8) is not evidence for non-linear mode coupling, for
the same reasons as given above.  We obtained the daughter mode (line
9) even when we eliminated all pre-existing waves by scrambling the
disc particles, and we have strong evidence that the new mode is a
linear instability caused by scattering at the ILR of the first mode.

The power spectrum in the bottom panel of Fig.~\ref{fig.edgevol}
strongly suggests a cascade of groove instabilities to ever higher
frequency, with corotation of each subsequent instability lying at the
ILR of the previous.  Since the initial outer edge mode had a low
growth rate, the inner disc had to wait for a succession of
destabilising grooves to be carved before each mode could develop.
This delay is all the more remarkable because the dynamical clock runs
faster at smaller radii.  The long absence of coherent disturbances in
the inner disc could therefore happen only if the single instability
the disc possesses is at its outer edge.

\subsection{Changes of angular symmetry}
Disturbance forces in the two sets of experiments described above were
each confined to a single sectoral harmonic: $m=2$ for the initial
groove mode (\S\ref{sec.groove}) and $m=3$ for the initial edge mode
(\S\ref{sec.edge}).  Furthermore, the simulations with scrambled or
spliced particle sets were also restricted to the same symmetry as in
the parent simulation.  Here we show that the features in action space
created by the first mode can also give rise to instabilities having
other rotational symmetries.

\begin{figure}
\includegraphics[width=.9\hsize,angle=0]{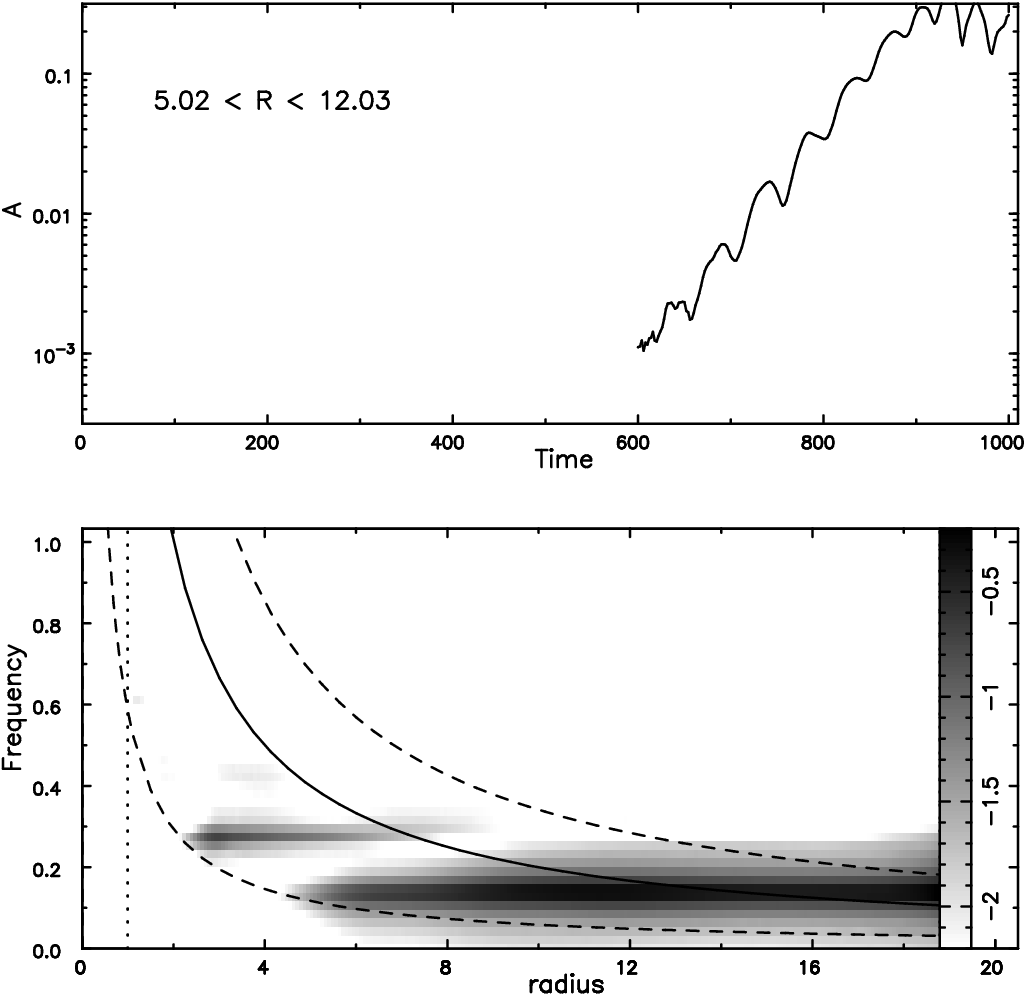}
\caption{The evolution of model ER2, the scrambled disc at $t=600$
  from model E after the $m=3$ edge mode had created the scattering
  features in Fig.~\ref{fig.act4887.600-0}, but in this case
  disturbance forces were restricted to $m=2$.  The top panel shows
  the evolution of the $m=2$ amplitude and the bottom panel presents
  the power spectrum over the time range $600 \leq t \leq 900$.}
\label{fig.run4892}
\end{figure}

\subsubsection{From 3 to 2}
The initial $m=3$ edge mode of model E presented in \S\ref{sec.edge}
scattered particles to produce the features shown in
Fig.~\ref{fig.act4887.600-0} by $t=600$.  We have already demonstrated
that the ILR scattering feature causes a new $m=3$ groove instability
that we identified both in the continued run, model E, and in a
separate simulation, model ER, started from a scrambled copy of the
particles at $t=600$.  Here we report another simulation, model ER2,
with the same scrambled copy of the particles at $t=600$, but with
disturbance forces restricted to $m=2$ instead.

The disturbance amplitude exponentiates rapidly, as shown in the top
panel of Fig.\ref{fig.run4892}, and the power spectrum in the bottom
panel reveals two simultaneous instabilities.  Our mode fitting
software finds two modes, a groove instability given in line 10 of
Table~\ref{tab.modes} and an edge mode having a frequency given in
line 11.

\begin{figure*}
\includegraphics[width=.9\hsize,angle=0]{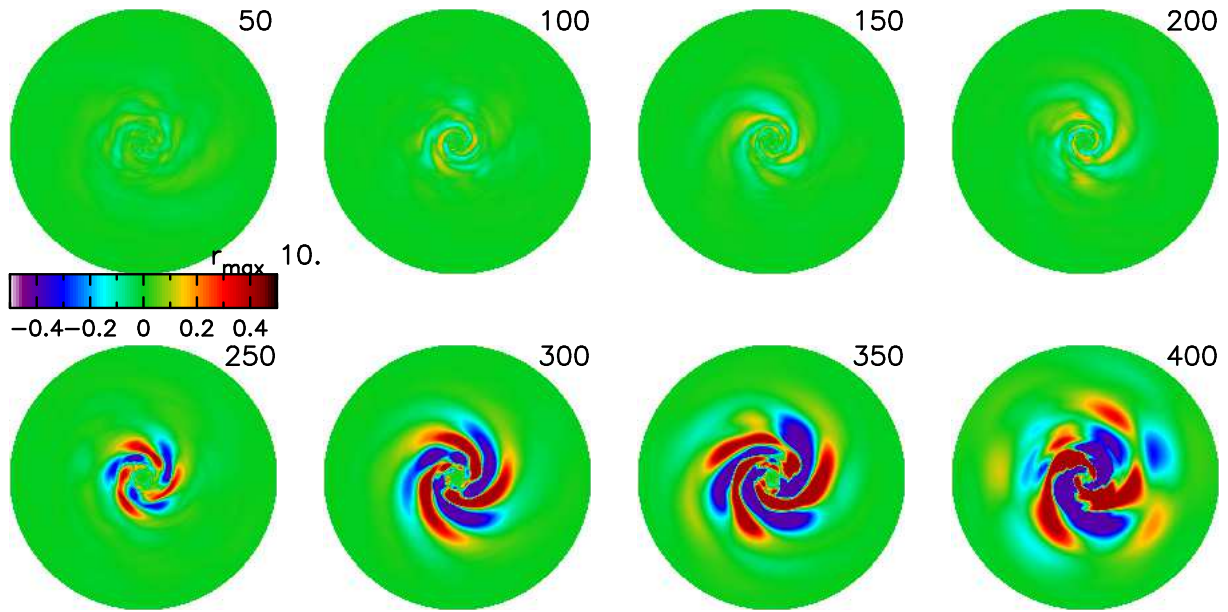}
\caption{Early part of the evolution of model GU, in which disturbance
  forces included $0 \leq m \leq 8$, except for $m=1$, terms.  The
  indicated colour scale shows the over- or under-density of
  disturbances that are scaled up by a factor 100.}
\label{fig.cntd4894}
\end{figure*}

As noted above (\S\ref{sec.e2}), the $m=3$ edge mode in model E was
still growing at $t=600$, and it is therefore hardly surprising to
find that the edge of this disc remains steep enough to excite a new
edge mode this time.  The new $m=2$ mode, line 11 of
Table~\ref{tab.modes}, has a higher growth rate than the $m=3$ edge
mode because the swing amplifier is at full strength in this case,
$X=2$, whereas $X=4/3$ for $m=3$.  Because the instability is more
vigorous, corotation is at the larger radius, as also indicated in the
Table.

The $m=2$ groove mode in model ER2 also has a higher growth rate, line
10, than at $m=3$ instability from the same file of particles (line 9
of Table~\ref{tab.modes}), consistent with more vigorous swing
amplification, and corotation is still farther from the groove created
by ILR scattering by the original $m=3$ edge mode at $R\simeq 6.9$, as
expected for a more vigorous and larger-scale groove instability
\citep{SK91}.

\subsubsection{From 2 to 3}
In this next case, model GS3$_{\rm OLR}$, we start from the spliced
set of particles shown in Fig.~\ref{fig.act4871.400-1}, but this time
restrict perturbation forces to the $m=3$ sectoral harmonic.  Our
estimated frequency of the dominant instability is given in line 12 of
Table~\ref{tab.modes}, and corotation is again almost exactly at the
location of the groove in action space, as for the modes in lines 2, 3
and 5.  Notice that the growth rate is a little lower than for the
corresponding $m=2$ case (line 5 of Table~\ref{tab.modes}), consistent
with expectations since swing amplification is less vigorous for $m=3$
than for $m=2$.

These last two experiments have demonstrated that a groove created by
Lindblad resonance scattering from a pattern of one sectoral harmonic
can excite an instability of a different $m$.

\begin{figure}
\includegraphics[width=.9\hsize,angle=0]{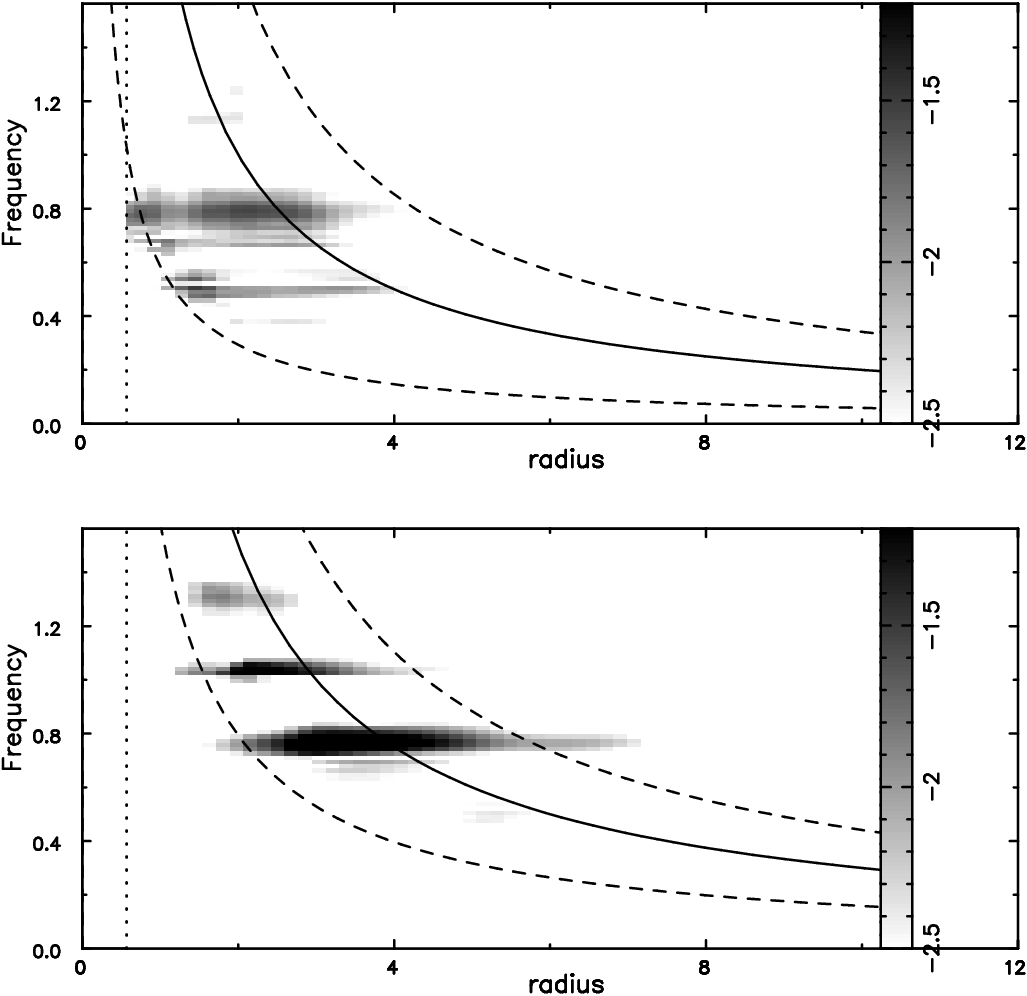}
\caption{Power spectra from model GU, \S\ref{sec.gen}, over the time
  range $0 \leq t \leq 400$.  The top panel is for $m=2$ and the
  bottom panel for $m=3$.}
\label{fig.run4894}
\end{figure}

\begin{figure}
\includegraphics[width=.9\hsize,angle=0]{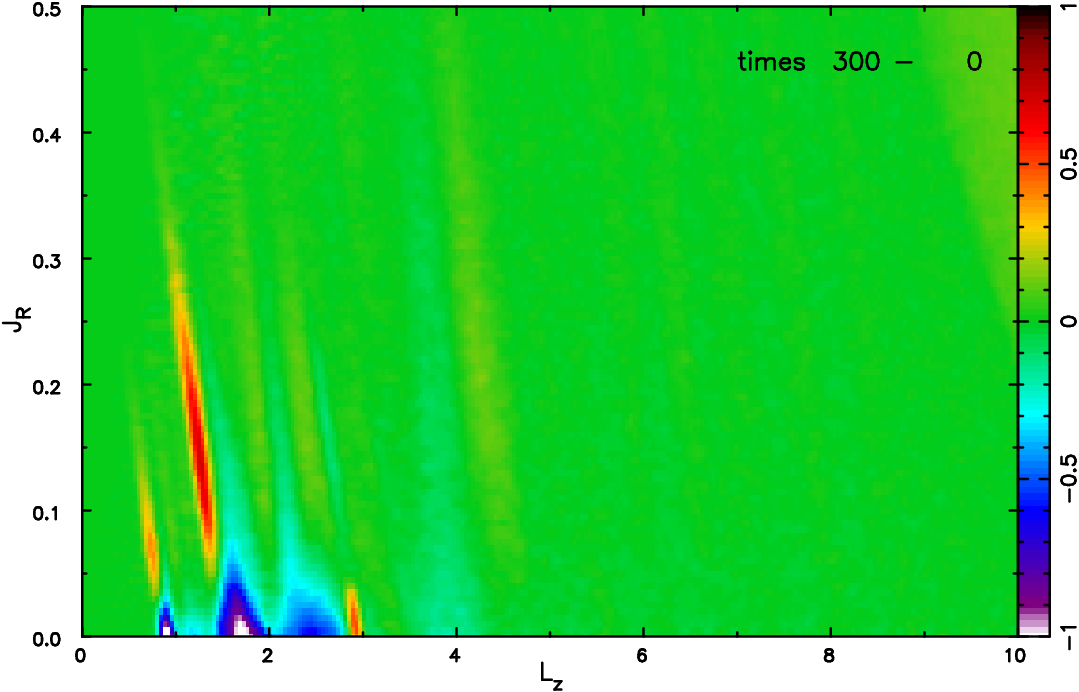}
\includegraphics[width=.9\hsize,angle=0]{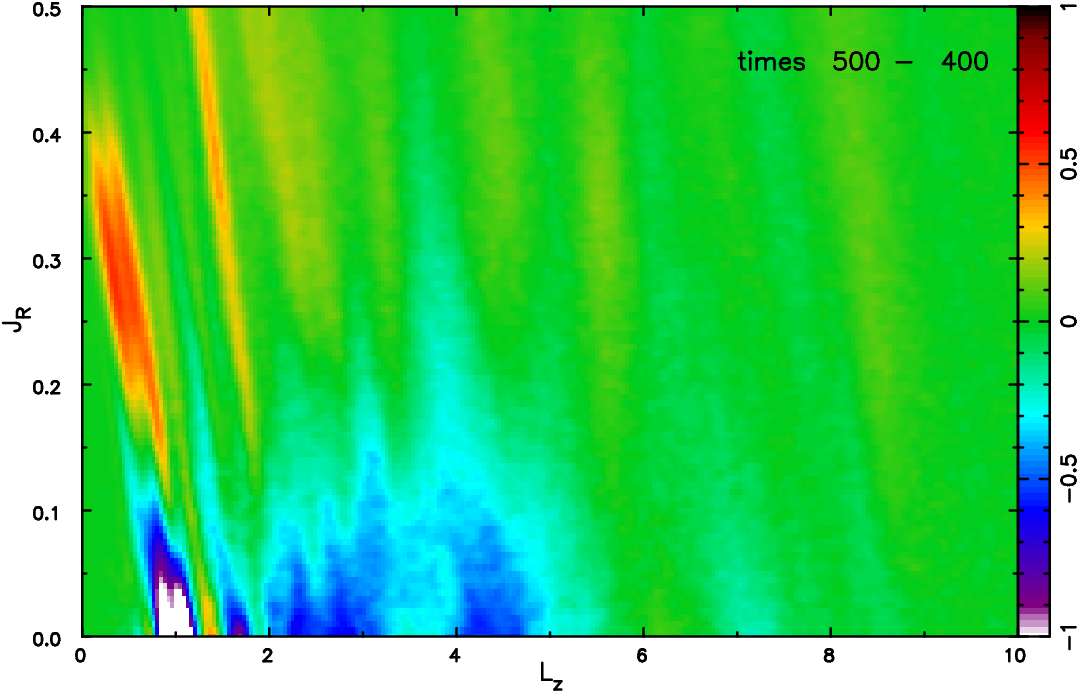}
\caption{Changes in action space of model GU (\S\ref{sec.gen}).  Top
  panel over the interval $t=0$ to $t=300$, bottom panel from $t=400$
  to $t=500$.  Note that the color scale has been compressed in
  comparison with other similar figures.}
\label{fig.act4894.300-0}
\end{figure}

\subsection{A very slightly more general case}
\label{sec.gen}
Perturbation forces in all the preceding simulations were restricted
to a single sectoral harmonic, which we now relax.  We start a
simulation, model GU, from the particle distribution used in the
groove mode case, shown in Fig.~\ref{fig.act4864.0-1}, which we know
from model G will excite at least a mild $m=2$ instability.  However,
perturbation forces in this new experiment now include $0 \leq m \leq
8$ terms, with the exception that $m=1$ is excluded to avoid possible
unbalanced forces from the fixed rigid halo component.

Multiple patterns develop over time, as shown in
Fig.~\ref{fig.cntd4894}, with perhaps the first feature to appear
being an $m=3$ mode. Fig.~\ref{fig.run4894} presents power spectra
from model GU over the same time interval for both $m=2$ and $m=3$,
revealing a number of coherent waves with pattern speeds independent
of radius.  A number of strong scattering features in action space are
visible at $t=300$ in Fig.~\ref{fig.act4894.300-0} (top).  The bottom
panel shows that changes become so intricate in the later evolution
that we are no longer able to isolate the effects of individual modes,
but there seems little doubt that the disc is being repeatedly
destabilized by scattering at resonances of successive disturbances,
each of which removes particles from near-circular orbits over narrow
ranges of $L_z$.

\section{Discussion}
\label{sec.discuss}
The simulations of \citet{SC14} manifested an apparently outwardly
propagating cascade of spiral modes to ever larger radii over time.
Although they presented evidence of resonance scattering and argued
that the changes to the DF excited subsequent modes, they did not
identify the mechanism.  Here we have presented compelling evidence
that the recurrence mechanism identified long ago by \citet{SL89} for
$m=4$ spirals a low-mass disc with a Keplerian rotation curve can, in
fact, operate in massive discs having flat rotation curves.  We
doubted that this would be viable for three principal reasons that, in
hindsight, no longer seem particularly cogent.
\begin{enumerate}[1]

\item Probably the strongest reason was that the radial extent of
  spiral modes, which are expected to extend from the ILR to the OLR
  \citep{LS64}, is much greater for $m=2$ waves in a disc having a
  flat rotation curve.  Both the lower $m$ value and the shape of the
  rotation curve cause the ratio $R_{\rm OLR}/R_{\rm ILR}$, which was
  1.67 for $m=4$ waves in a Kepler potential, to rise to $\sim 5.8$ for
  $m=2$ patterns in a flat rotation curve.  If the patterns were
  required not to overlap, it would be hard to fit many into any disc
  of reasonable extent.  But in fact, for a cascade of outwardly
  propagating $m=2$ modes, successive disturbances have radii in the
  ratio $R_{\rm OLR}/R_{\rm CR} \simeq 1.7$.  Even for an inwardly
  propagating $m=2$ cascade initiated by an edge mode, say, the ratio
  is $R_{\rm CR}/R_{\rm ILR} \simeq 3.4$.  These smaller ratios, which
  also allow extensive overlap of successive modes, would be even
  further reduced for larger values of $m$.

\item It is well known that $m=2$ spiral disturbances have much
  greater amplitude inside corotation than outside -- examples include
  the famous dust-to-ashes figure from \citet{To81}, as well as the
  power spectra and modes presented here, \eg\ the middle panel of
  Fig.~\ref{fig.grevol}.  However, the fitted mode in that case
  (Fig.~\ref{fig.mode4864}) does have a weak spiral that extends past
  the OLR circle.  This general disparity suggested that scattering at
  the OLR would not be as important as at the ILR.  Since self-excited
  modes in a disc must conserve angular momentum (there is no
  externally applied torque), the gains and losses must balance and
  the angular momentum emitted at the ILR must be absorbed at other
  resonances.  With the exception of edge modes, where corotation
  takes up almost all the angular momentum lost at the ILR
  (Fig.~\ref{fig.act4887.600-0}), the OLR must absorb a significant
  amount.

\item The third reason we expected that OLR scattering would be less
  consequential than at the ILR has already been stated.  The close
  alignment of the ILR locus and the scattering vector causes
  particles to stay on resonance as they are scattered allowing large
  changes to build up.  This does not happen at the OLR, where
  scattered particles are quickly moved off resonance and the angular
  momentum gain must therefore be shared among many more particles.
  Again this is true and consistent with our results presented in
  Fig.~\ref{fig.act4864.400-0}, which led us to doubt that the feeble
  OLR scattering seen there could excite the second mode.  However, it
  seems that the depopulation of the low-$J_R$ region, as many
  particles are shifted away from near-circular orbits, is the
  dominant source of excitation of the next mode.
\end{enumerate}
These arguments make it clear that we were wrong to doubt the
possibility of a cascade of groove instabilities, running either
radially outward or inward, could occur in a massive disc having a
flat rotation curve.  The evidence presented in \S\ref{sec.results}
demonstrates that cascades of groove modes, with each causing changes
to excite the next, arise naturally.

\section{Conclusions}
\label{sec.concl}
In this paper, we have presented a new part of our picture that spiral
patterns result from true instabilities in galaxy discs.  We have
demonstrated that scattering at either Lindblad resonance by any one
wave carves a new groove in the disc that excites a further mode.  We
presented examples of both inwardly and outwardly propagating cascades
of instabilities in simulations restricted to a single sectoral
harmonic.  We also showed that a groove created by one wave of a
certain angular symmetry can excite a mode of different angular
symmetry.  Thus, in more general simulations that include disturbance
forces from multiple sectoral harmonics, such as are inevitable when
tree codes are employed, it quickly becomes very difficult to follow
the causal chain from one wave to the next.

The mirror mode mechanism, proposed by \citet{SC14}, seemed to be
responsible for the first true mode in models 50 of \citet{Se12}, as
evidenced by Fig.~\ref{fig.mode4169} shown here.  However, we suspect
it is important only in models that lack strong instabilities.  Once a
recurrent cycle of groove modes becomes established, mild variations
of impedance to cause partial reflections of travelling waves must
become physically less important as scattering features in phase space
increase in both the strength and number -- see \eg\ the bottom panel
of our Fig.~\ref{fig.act4894.300-0} and Fig.~6 of \citet{SC14}, 
where 3D motion was allowed.  The recurrent cycle of groove modes
reported here is then far more likely to be the main driver of
evolution.

The additional evidence we have presented here strengthens our
argument that spirals in simulations are manifestations of global
instabilities.  The modes in our simulations are each detectable as
coherent waves of fixed frequency for a number of rotations, but the
visual pattern of spirals changes on a much shorter time scale.  The
superposition of two or more modes inevitably leads to the apparently
shearing spiral features that are widely reported.  But only unbounded
growth of true instabilities in the linear regime can account for the
final amplitudes being independent of the number of particles, as we
have repeatedly shown in our previous work \citep{Se11, Se12, SC14}.
None of the many papers arguing for the fundamental nature of shearing
features has so far presented any mechanism that could deliver
unbounded growth.  Swing amplification and wakes, which are invoked in
these papers, undoubtedly enhance collective features, but only by a
fixed factor in linear theory \citep{JT66, To81}.  \citet{DVH}
describe non-linear consequences of responses to the mass clumps they
introduced into their disc models, which do appear to cause more
persisting non-axisymmetric structures, but it is unclear as yet
whether they could account for indefinite growth.  The non-linear
scattering at resonances, that is a crucial part of our picture,
readily excites new instabilities, leading to a robust cycle of
recurring modes that accounts for the kind of behaviour that is almost
universally observed in simulations of disc galaxies.

Naturally, resonant scattering causes irreversible changes, and the
consequent secular increase in random motion would, if unchecked,
degrade the ability of the disc to support continuing collective
instabilities, and spiral activity must fade.  But \citet{SC84} were
the first to show that allowing for dissipation and star formation in
a reasonable gas fraction could allow spiral activity to continue
``indefinitely,'' an idea that has been supported in much subsequent
work.

In their analysis of the local {\it Gaia\/} data, \citet{Se19} argued
that an action-space representation of local phase space revealed
evidence for Lindblad resonance scattering, that was not all
attributable to the bar \citep{Mo19}.  This evidence suggests that the
Milky Way has supported global spiral modes of the type we describe in
our simulations, and we are reasonably confident that spirals in the
old stellar discs of many other galaxies also result from such
instabilities.

\section*{Acknowledgements}
We thank an anonymous referee for thoughtful comments on the
paper, and also Elena D'Onghia and especially Daisuke Kawata for
helpful conversations during the Gaia19 program at KITP, where this
paper was advanced significantly.  KITP is supported in part by NSF
grant PHY-1748958. JAS also acknowledges the hospitality of Steward
Observatory.


\bsp	
\label{lastpage}

\begin{thebibliography}{99}
\def\skip#1{ \etal\ }
\def\PhD{PhD thesis.}
\def\rmp{Rev. Mod. Phys.}
\def\rpp{Rep. Prog. Phys.}

\bibitem[\protect\citeauthoryear{Agertz \etal}{2011}]{Ag11}
Agertz, O., Teyssier, R. \& Moore, B. 2011, \mnras, {\bf 410}, 1391

\bibitem[\protect\citeauthoryear{Athanassoula \etal}{1987}]{ABP}
Athanassoula, E., Bosma, A. \& Papaioannou, S. 1987, \aap, {\bf 179}, 23

\bibitem[\protect\citeauthoryear{Aumer \etal}{2016}]{Au16}
Aumer, M., Binney, J. \& Sch\"onrich, R. 2016, \mnras, {\bf 459}, 3326

\bibitem[\protect\citeauthoryear{Baba}{2015}]{Baba15}
Baba, J. 2015, \mnras, {\bf 454}, 2954

\bibitem[\protect\citeauthoryear{Baba \etal}{2013}]{Baba13}
Baba, J., Saitoh, T. R. \& Wada, K. 2013, \apj, {\bf 763}, 46

\bibitem[\protect\citeauthoryear{Bertin \& Lin}{1996}]{BL96}
Bertin, G. \& Lin, C. C. 1996, {\it Spiral Structure in Galaxies\/} (Cambridge, MA: The MIT Press)

\bibitem[\protect\citeauthoryear{Bertin \etal}{1989}]{BLLT}
Bertin, G., Lin, C. C., Lowe, S. A. \& Thurstans, R. P. 1989, \apj, {\bf 338}, 104

\bibitem[\protect\citeauthoryear{Binney}{2019}]{Bi19}
Binney, J. 2019, \mnras, submitted (arXiv:1906.11696)

\bibitem[\protect\citeauthoryear{Binney \& Tremaine}{2008}]{BT08}
Binney J. \& Tremaine S. 2008, \textit{Galactic Dynamics} 2nd ed. (Princeton University Press, Princeton NJ) (BT08)

\bibitem[\protect\citeauthoryear{Carlberg \& Sellwood}{1985}]{CS85}
Carlberg, R. G. \& Sellwood, J. A. 1985, \apj, {\bf 292}, 79

\bibitem[\protect\citeauthoryear{Davis \etal}{2012}]{Davi12}
Davis, B. L., Berrier, J. C., Shields, D. W.,\skip{ Kennefick, J., Kennefick, D., Seigar, M. S., Lacy, C. H. S. \& Puerari, I.} 2012, \apjs, {\bf 199}, 33

\bibitem[\protect\citeauthoryear{Debattista \& Sellwood}{2000}]{DS00}
Debattista, V. P. \& Sellwood, J. A. 2000, \apj, {\bf 543}, 704

\bibitem[\protect\citeauthoryear{De Rijcke \etal}{2019}]{DFP19}
De Rijcke, S., Fouvry, J-B. \& Pichon, C. 2019, \mnras, {\bf 484}, 3198

\bibitem[\protect\citeauthoryear{Dobbs \& Baba}{2014}]{DB14}
Dobbs, C. \& Baba, J. 2014, PASA, {\bf 31}, 35

\bibitem[\protect\citeauthoryear{D'Onghia \etal}{2013}]{DVH}
D'Onghia, E., Vogelsberger, M. \& Hernquist, L. 2013, \apj, {\bf 766}, 34

\bibitem[\protect\citeauthoryear{Earn \& Sellwood}{1995}]{ES95}
Earn, D. J. D. \& Sellwood, J. A. 1995, \apj, {\bf 451}, 533

\bibitem[\protect\citeauthoryear{Fouvry \etal}{2015}]{FBP15}
Fouvry, J-B., Binney, J. \& Pichon, C. 2015, \apj, {\bf 806}, 117

\bibitem[\protect\citeauthoryear{Fouvry \& Pichon}{2015}]{FP15}
Fouvry, J-B. \& Pichon, C. 2015, \mnras, {\bf 449}, 1928

\bibitem[\protect\citeauthoryear{Gaia collaboration: Katz \etal}{2018}]{Gaia2}
Gaia collaboration: Katz, D., Antoja, T., Romero-G\'o, M., \etal\ 2018, \aap, {\bf 616A}, 11 

\bibitem[\protect\citeauthoryear{Grand \etal}{2012a}]{Gran12a}
Grand, R. J. J., Kawata, D. \& Cropper, M.  2012a, \mnras, {\bf 421}, 1529

\bibitem[\protect\citeauthoryear{Grand \etal}{2012b}]{Gran12b}
Grand, R. J. J., Kawata, D. \& Cropper, M.  2012b, \mnras, {\bf 426}, 167

\bibitem[\protect\citeauthoryear{Grand \etal}{2013}]{Gran13}
Grand, R. J. J., Kawata, D. \& Cropper, M.  2013, \aap, {\bf 553}A, 77

\bibitem[\protect\citeauthoryear{Goldreich \& Lynden-Bell}{1965}]{GLB}
Goldreich, P. \& Lynden-Bell, D. 1965, \mnras, {\bf 130}, 97

\bibitem[\protect\citeauthoryear{Hart \etal}{2016}]{Hart16}
Hart, R. E., Bamford, S. P., Willett, K. W.,\skip{ Masters, K. L., Cardamone, C., Lintott, C. J., Mackay, R. J., Nichol, R. C., Rosslowe, C, K., Simmons, B. D. \& Smethurst, R. J.} 2016, \mnras, {\bf 461}, 3663

\bibitem[\protect\citeauthoryear{Hohl}{1971}]{Ho71}
Hohl, F. 1971, \apj, {\bf 168}, 343

\bibitem[\protect\citeauthoryear{Hunt \etal}{2018}]{Hu18}
Hunt, J. A. S., Hong, J., Bovy, J., Kawata, D. \& Grand, R. J. J. 2018, \mnras, {\bf 481}, 3794 

\bibitem[\protect\citeauthoryear{Julian \& Toomre}{1966}]{JT66}
Julian, W. H. \& Toomre, A. 1966, \apj, {\bf 146}, 810

\bibitem[\protect\citeauthoryear{Kalnajs}{1971}]{Ka71}
Kalnajs, A. J. 1971, \apj, {\bf 166}, 275

\bibitem[\protect\citeauthoryear{Kalnajs}{1973}]{Ka73}
Kalnajs, A. J. 1973, {\it Proc. Astron. Soc. Australia}, {\bf 2}, 174

\bibitem[\protect\citeauthoryear{Kalnajs}{1978}]{Ka78}
Kalnajs, A. J. 1978, in IAU Symposium {\bf 77} {\it Structure and Properties of Nearby Galaxies} eds.\ E. M. Berkhuisjen \& R. Wielebinski (Dordrecht:Reidel) p.~113

\bibitem[\protect\citeauthoryear{Kawata \etal}{2014}]{Ka14}
Kawata, D., Hunt, Jason A. S., Grand, R. J. J., Pasetto, S. \& Cropper, M. 2014, \mnras, {\bf 443}, 2757

\bibitem[\protect\citeauthoryear{Kumamoto \& Noguchi}{2016}]{KN16}
Kumamoto, J. \& Noguchi, M. 2016, \apj, {\bf 822}, 110

\bibitem[\protect\citeauthoryear{Lin \& Shu}{1964}]{LS64}
Lin, C. C. \& Shu, F. H. 1964, \apj, {\bf 140}, 646

\bibitem[\protect\citeauthoryear{Lynden-Bell \& Kalnajs}{1972}]{LBK}
Lynden-Bell, D. \& Kalnajs, A. J. 1972, \mnras, {\bf 157}, 1

\bibitem[\protect\citeauthoryear{Mark}{1974}]{Ma74}
Mark, J. W-K. 1974, \apj, {\bf 193}, 539

\bibitem[\protect\citeauthoryear{Mark}{1977}]{Ma77}
Mark, J. W-K. 1977, \apj, {\bf 212}, 645

\bibitem[\protect\citeauthoryear{Michikoshi \& Kokubo}{2018}]{MK18}
Michikoshi, S. \& Kokubo, E. 2018, \mnras, {\bf 481}, 185

\bibitem[\protect\citeauthoryear{Monari \etal}{2018}]{Mo19}
Monari, G., Famaey, B., Siebert, A., Wegg, C. \& Gerhard, O. 2019, \aap, {\bf 626}A, 41

\bibitem[\protect\citeauthoryear{Oort}{1962}]{Oo62}
Oort, J. H. 1962, in {\it Interstellar Matter in Galaxies}, ed.\ L. Woltjer (New York: Benjamin), p.~234

\bibitem[\protect\citeauthoryear{Papaloizou \& Lin}{1989}]{PL89}
Papaloizou, J. C. B. \& Lin, D. N. C. 1989, \apj, {\bf 344}, 645

\bibitem[\protect\citeauthoryear{Quillen \etal}{2011}]{Qu11}
Quillen, A. C., Dougherty, J., Bagley, M. B., Minchev, I. \& Comparetta, J. 2011, \mnras, {\bf 417}, 762

\bibitem[\protect\citeauthoryear{Roca-F\`abrega \etal}{2013}]{Roca13}
Roca-F\`abrega, S., Valenzuela, O., Figueras, F.,\skip{ Romero-G\'omez, M., Vel\'azquez, H., Antoja, T. \& Pichardo, B.} 2013, \mnras, {\bf 432}, 2878

\bibitem[\protect\citeauthoryear{Ro\v skar \etal}{2008}]{Ro08}
Ro\v skar, R., Debattista, V. P., Quinn, T. R., Stinson, G. S. \& Wadsley, J. 2008, \apjl, {\bf 684}, L79

\bibitem[\protect\citeauthoryear{Sellwood}{1989}]{Se89}
Sellwood, J. A. 1989, in {\it Dynamics of Astrophysical Discs}, ed.\ J. A. Sellwood (Cambridge: Cambridge University Press) p.~155

\bibitem[\protect\citeauthoryear{Sellwood}{2011}]{Se11}
Sellwood, J. A. 2011, \mnras, {\bf 410}, 1637

\bibitem[\protect\citeauthoryear{Sellwood}{2012}]{Se12}
Sellwood, J. A. 2012, \apj, {\bf 751}, 44

\bibitem[\protect\citeauthoryear{Sellwood}{2014a}]{Se14a}
Sellwood, J. A. 2014a, \rmp, {\bf 86}, 1

\bibitem[\protect\citeauthoryear{Sellwood}{2014b}]{Se14b}
Sellwood, J. A. 2014b, arXiv:1406.6606 (on-line manual: \hfil\break {\tt http://www.physics.rutgers.edu/$\sim$sellwood/manual.pdf})

\bibitem[\protect\citeauthoryear{Sellwood \& Athanassoula}{1986}]{SA86}
Sellwood, J. A. \& Athanassoula, E. 1986, \mnras, {\bf 221}, 195

\bibitem[\protect\citeauthoryear{Sellwood \& Binney}{2002}]{SB02}
Sellwood, J. A. \& Binney, J. J. 2002, \mnras, {\bf 336}, 785

\bibitem[\protect\citeauthoryear{Sellwood \& Carlberg}{1984}]{SC84}
Sellwood, J. A. \& Carlberg, R. G. 1984, \apj, {\bf 282}, 61

\bibitem[\protect\citeauthoryear{Sellwood \& Carlberg}{2014}]{SC14}
Sellwood, J. A. \& Carlberg, R. G. 2014, \apj, {\bf 785}, 137

\bibitem[\protect\citeauthoryear{Sellwood \& Kahn}{1991}]{SK91}
Sellwood, J. A. \& Kahn, F. D. 1991, \mnras, {\bf 250}, 278

\bibitem[\protect\citeauthoryear{Sellwood \& Lin}{1989}]{SL89}
Sellwood, J. A. \& Lin, D. N. C. 1989, \mnras, {\bf 240}, 991

\bibitem[\protect\citeauthoryear{Sellwood \etal}{2019}]{Se19}
Sellwood, J. A., Trick, W. H., Carlberg, R. G., Coronado, J. \& Rix, H-W. 2019, \mnras, {\bf 484}, 3154

\bibitem[\protect\citeauthoryear{Shu}{2016}]{Sh16}
Shu, F. H. 2016, \araa, {\bf 54}, 667

\bibitem[\protect\citeauthoryear{Sridhar}{2019}]{Sr19}
Sridhar, S. 2019, arXiv:1906.08655

\bibitem[\protect\citeauthoryear{Tagger \etal}{1987}]{Ta87}
Tagger, M., Sygnet, J. F., Athanassoula, E. \& Pellat, R. 1987, \apjl, {\bf 318}, 43

\bibitem[\protect\citeauthoryear{Toomre}{1969}]{To69}
Toomre, A. 1969, \apj, {\bf 158}, 899

\bibitem[\protect\citeauthoryear{Toomre}{1977}]{To77}
Toomre, A. 1977, \araa, {\bf 15}, 437

\bibitem[\protect\citeauthoryear{Toomre}{1981}]{To81}
Toomre, A. 1981, In ''The Structure and Evolution of Normal Galaxies'', eds.~S. M. Fall \& D. Lynden-Bell (Cambridge, Cambridge Univ. Press) p.~111

\bibitem[\protect\citeauthoryear{Toomre}{1989}]{To89}
Toomre, A. 1989, in {\it Dynamics of Astrophysicsl Discs}, ed.\ J. A. Sellwood (Cambridge: Cambridge University Press) p.~153
  
\bibitem[\protect\citeauthoryear{Toomre}{1990}]{To90}
Toomre, A. 1990, in {\it Dynamics \& Interactions of Galaxies}, ed.\ R. Wielen (Berlin, Heidelberg: Springer-Verlag), p.~292

\bibitem[\protect\citeauthoryear{Toomre \& Kalnajs}{1991}]{TK91}
Toomre, A. \& Kalnajs, A. J. 1991, in {\it Dynamics of Disc Galaxies}, ed.\ B. Sundelius (Gothenburg: G\"oteborgs University) p.~341

\bibitem[\protect\citeauthoryear{Yu \etal}{2018}]{Yu18}
Yu, S-Y., Ho, L. C., Barth, A. J. \& Li, Z-Y. 2018, \apj, {\bf 862}, 13

\bibitem[\protect\citeauthoryear{Zang}{1976}]{Za76}
Zang, T. A, 1976, \PhD, MIT


\end{thebibliography}
\end{document}